\documentclass[%
reprint,
 superscriptaddress,
 amsmath,amssymb,
 aps,
prb,
floatfix,
longbibliography, 
notitlepage]{revtex4-1}

\usepackage[usenames,dvipsnames]{xcolor}
\usepackage{graphicx}
\usepackage{dcolumn}
\usepackage{bm}
\usepackage[normalem]{ulem}
\usepackage{hyperref}
\hypersetup{colorlinks = true, citecolor = blue, breaklinks = true}
\usepackage{textcomp}
\usepackage{multirow}
\usepackage{adjustbox}
\usepackage{xfrac}
\usepackage[version=4]{mhchem}
\usepackage[resetlabels]{multibib}
\newcites{SI}{References}

\begin{document}

\title{Importance of surface oxygen vacancies for ultrafast hot carrier relaxation and transport in \ce{Cu2O}}

\affiliation{%
Department of Chemistry and Biochemistry, University of Bern, Freiestrasse 3, CH-3012 Bern, Switzerland 
}%
\affiliation{%
National Centre for Computational Design and Discovery of Novel Materials (MARVEL), Switzerland
}%
\affiliation{%
Department of Physics, University of Zurich, Winterthurerstrasse 190, CH-8057 Zurich, Switzerland
}%

\author{Chiara Ricca}
\affiliation{%
Department of Chemistry and Biochemistry, University of Bern, Freiestrasse 3, CH-3012 Bern, Switzerland 
}%
\affiliation{%
National Centre for Computational Design and Discovery of Novel Materials (MARVEL), Switzerland
}%

\author{Lisa Grad}
\affiliation{%
Department of Physics, University of Zurich, Winterthurerstrasse 190, CH-8057 Zurich, Switzerland
}%

\author{Matthias Hengsberger}
\affiliation{%
Department of Physics, University of Zurich, Winterthurerstrasse 190, CH-8057 Zurich, Switzerland
}%

\author{ J{\"u}rg Osterwalder}
\affiliation{%
Department of Physics, University of Zurich, Winterthurerstrasse 190, CH-8057 Zurich, Switzerland
}%

\author{Ulrich Aschauer}
\email{ulrich.aschauer@dcb.unibe.ch}
\affiliation{%
Department of Chemistry and Biochemistry, University of Bern, Freiestrasse 3, CH-3012 Bern, Switzerland 
}%
\affiliation{%
National Centre for Computational Design and Discovery of Novel Materials (MARVEL), Switzerland
}%

\date{\today}

\begin{abstract}
\ce{Cu2O} has appealing properties as an electrode for photo-electrochemical water splitting, yet its practical performance is severely limited by inefficient charge extraction at the interface. Using hybrid DFT calculations, we investigate carrier capture processes by oxygen vacancies (V$_\mathrm{O}$) in the experimentally observed ($\sqrt{3} \times \sqrt{3}$)R30$^{\circ}$ reconstruction of the dominant (111) surface. Our results show that these V$_\mathrm{O}$ are doubly ionized and that associated defects states strongly suppress electron transport. In particular, the excited electronic state of a singly charged V$_\mathrm{O}$ plays a crucial role in the non-radiative electron capture process with a capture coefficient of about 10$^{-9}$~cm$^3$/s and a lifetime of 0.04~ps, explaining the experimentally observed ultrafast carrier relaxation. These results highlight that engineering the surface V$_\mathrm{O}$ chemistry will be a crucial step in optimizing \ce{Cu2O} for photoelectrode applications.
\end{abstract}

\maketitle

\section{\label{sec:intro}Introduction}

Cuprous oxide (\ce{Cu2O}) is a promising material for a variety of industrial applications due to its small direct band gap, its high absorbance, the abundance and non-toxicity of its constituent elements and the large flexibility and low cost of \ce{Cu2O}-based thin-film preparation methods~\cite{Meyer2012}. In particular, cuprous oxide has recently attracted much attention as an electrode material for photo-electrochemical water splitting with efficient light absorption, high positive onset voltage, and high photocurrent density~\cite{Niu2018,Pan2018}. However, the electrode performance is highly sensitive to the presence of defect states generally appearing within the semiconductor band gap~\cite{Borgwardt2019,Grad2020}. Such defect states may trap excited charge carriers resulting in a reduction of the generated photovoltage and photocurrent.

In bulk \ce{Cu2O} both copper and oxygen vacancies occur, copper vacancies being more abundant and generally leading to \textit{p}-type doping. The (111) facet is the experimentally most studied \ce{Cu2O} surface and two ordered surface structures were identified. The first corresponds to the ideal O-terminated and stoichiometric surface with ($1 \times 1$) periodicity. The second is a reconstructed ($\sqrt{3} \times \sqrt{3}$)R30$^{\circ}$ surface with only \sfrac{2}{3} of the O ions in the outermost surface layer~\cite{SchulzCox1991, Onsten2009, Onsten2013}. The exact termination, stoichiometry and atomic structure of the reconstructed surface have been studied by experiment~\cite{SchulzCox1991, Onsten2009, Onsten2013} and theory~\cite{SUN2008, Bendavid2013JPCB, Zhang2018, Yu2018, Gloystein2020}, associating it most likely with a \sfrac{1}{3} monolayer of ordered oxygen vacancies (V$_\mathrm{O}$)~\cite{SchulzCox1991, Onsten2009, Onsten2013}.

Recently, it was suggested that these surface V$_\mathrm{O}$ dominantly affect the photocatalytic performance in \ce{Cu2O}-based materials under oxygen poor conditions~\cite{Grad2020}. Comparing electron dynamics at stoichiometric and oxygen deficient \ce{Cu2O}-(111) surfaces observed via time-resolved two-photon photoemission (2PPE) indicates that surface rather than bulk defects limit the photovoltage. Excited electrons can drift within the conduction band to the stoichiometric \ce{Cu2O}-(111) surface where they create an energetic electron population that decays with a characteristic time of 10 picoseconds. On the oxygen deficient ($\sqrt{3} \times \sqrt{3}$)R30$^{\circ}$-(111) surface, however, no indication of electrons occupying the conduction band was found, but instead electrons are trapped, within 1 picosecond, by low-lying defect states with no further noticeable change of the electron population over a timescale of hundreds of picoseconds. This indicates that V$_\mathrm{O}$ associated with the reconstruction likely trap carriers.

Unfortunately, previous experimental~\cite{SchulzCox1991, Onsten2009, Onsten2013, KOIRALA201465} and theoretical~\cite{Nolan2006, Raebiger2007, Nolan008, ScanlonJPCL2010, ScanlonJPCL2010, Huang2016} work reports contradictory results for V$_\mathrm{O}$ in \ce{Cu2O}, especially in terms of the thermodynamic stability of different charge states, the position of the corresponding defect levels in the band gap, and consequently their ability to act as efficient carrier traps (see supporting information (SI) Section~\ref{sec:defbulk}). In addition, these theoretical studies are restricted to the bulk and the experimentally relevant carrier capture at surface defects has never been computationally studied.

Here we investigate the stability and electronic properties of V$_\mathrm{O}$, and the electron and hole trapping processes by these defects at the \ce{Cu2O} surface by combining density functional theory (DFT) calculations and photoemission experiments. Accurately computing the electronic properties of \ce{Cu2O} is a challenging task, due to the peculiar properties of this oxide: a mainly ionic semiconductor with closed shell \ce{Cu+} ions and a higher coordination of O than Cu atoms, which leads to large cohesive energies. Semi-local DFT fails to predict the semiconducting character~\cite{Nolan008, Meyer2012}, DFT+$U$ approaches also failing to open the band gap due to the fully occupied Cu-$3d^{10}$ states. All our calculations have hence been performed using hybrid functionals, which can successfully reproduce the band structure of \ce{Cu2O}~\cite{ScanlonPRL2009, ScanlonJPCL2010}, but are rather expensive for applications in solid-state chemistry, especially with planewave-based codes and for surface/defect calculations that require large supercells. By comparing computed defect levels and electron capture coefficients with experimental results, we reach a deep understanding of the carrier dynamics and factors limiting photoelectrode performance. In particular, our results demonstrate that the defect states associated with the reconstructed \ce{Cu2O}-(111) surface are due to an ordered arrangement of doubly ionized V$_\mathrm{O}$, which effectively trap electrons and strongly suppress electron transport. This indicates that engineering the surface defect chemistry of \ce{Cu2O}-based photoelectrodes is crucial to enhance their performance.

\section{\label{sec:results}Results and Discussion}

\subsection{($\sqrt{3} \times \sqrt{3}$)R30$^{\circ}$- (111) \ce{Cu2O} surface}

\begin{figure}
 \centering
 \includegraphics[width=\columnwidth]{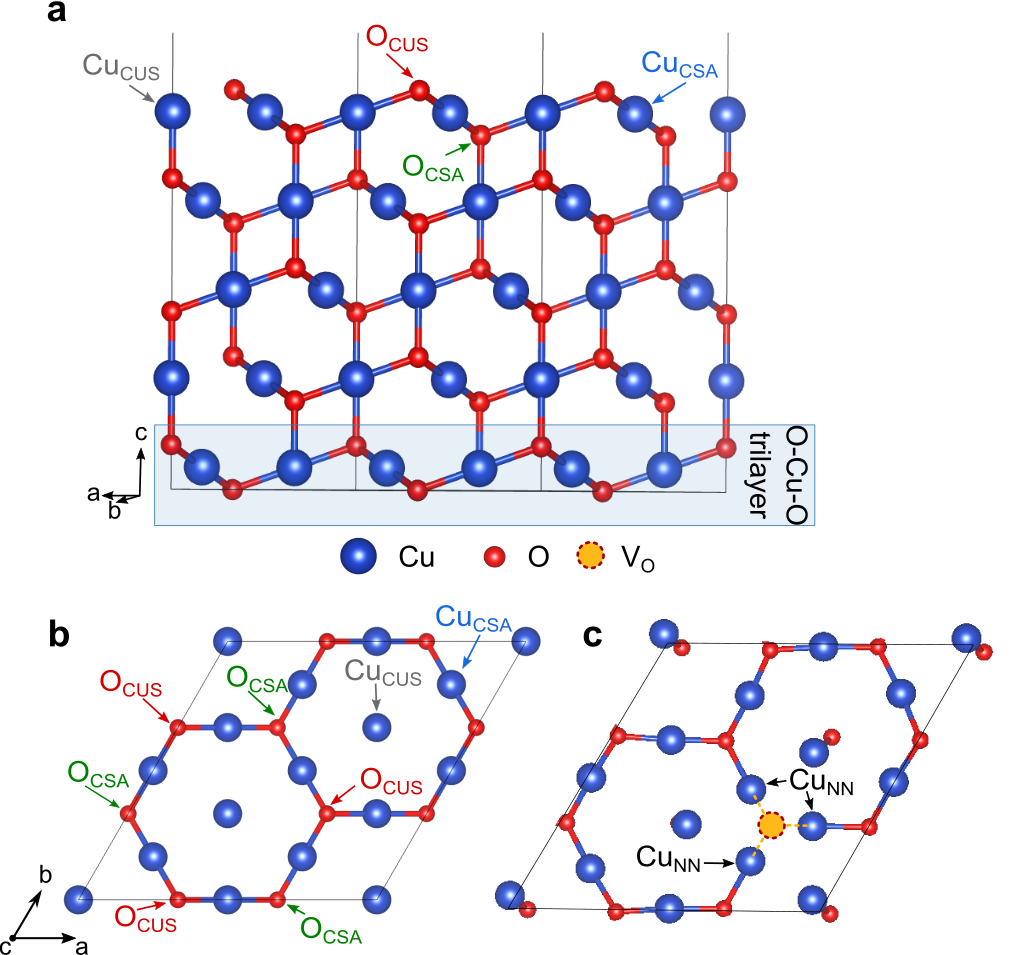}
 \caption{a) Lateral and b) top view of the stoichiometric, unreconstructed ($1 \times 1$)-(111) \ce{Cu2O} slab model within the ($\sqrt{3} \times \sqrt{3}$)R30$^{\circ}$ supercell. O$_\mathrm{CUS}$, Cu$_\mathrm{CUS}$ are the coordinatively unsaturated O and Cu surface sites, respectively, while O$_\mathrm{CSA}$, and O$_\mathrm{CSA}$ indicate the coordinatively saturated O and Cu atoms. c) Top view of the reconstructed ($\sqrt{3} \times \sqrt{3}$)R30$^{\circ}$-(111) \ce{Cu2O} slab model with one oxygen vacancy (V$_\mathrm{O}$). Cu$_\mathrm{NN}$ are Cu sites adjacent to V$_\mathrm{O}$.}
\label{fig:stoichsurface}
\end{figure}

 Figure~\ref{fig:stoichsurface} illustrates the slab model used to simulate the unreconstructed ($1 \times 1$) and reconstructed ($\sqrt{3} \times \sqrt{3}$)R30$^{\circ}$-(111) \ce{Cu2O} surfaces within the ($\sqrt{3} \times \sqrt{3}$)R30$^{\circ}$ supercell, where the $c$-axis is perpendicular to the surface plane. The unreconstructed surface consists of O--Cu--O trilayers, with each Cu layer sandwiched between two O layers. The topmost trilayer in the unreconstructed supercell contains four types of atoms: 3 coordinatively unsaturated O atoms (O$_\mathrm{CUS}$) in the first layer, 3 coordinatively unsaturated (Cu$_\mathrm{CUS}$) and 9 coordinatively saturated Cu atoms (Cu$_\mathrm{CSA}$) in the second, and 3 coordinatively saturated O atoms (O$_\mathrm{CSA}$) in the third layer. Experiment suggests that the reconstructed oxygen-deficient surface, which we focus on in this work, is associated with a \sfrac{1}{3} monolayer of charged surface V$_\mathrm{O_{CUS}}$ forming ordered structures due to mutual electrostatic repulsion~\cite{SchulzCox1991, Onsten2009, Onsten2013}.

Surface energies computed for the ideal stoichiometric (0.831~J/m$^2$) and defective (0.919/0.931~J/m$^2$ under O-poor/rich conditions) surface are slightly higher (by about 0.2~J/m$^2$ and 0.1~J/m$^2$ respectively) compared to previous semi-local DFT or DFT+$U$ results~\cite{Bendavid2013JPCB}. The qualitative picture is, however, in agreement, indicating the ideal surface to be slightly more stable than the reconstructed surface, regardless of the O chemical potential. This is also in line with the experimental conditions required to obtain the two surface structures: the ideal surface is generally obtained using milder conditions for the ion bombardment and with high-temperature annealing while the reconstructed surface is obtained for ion bombardment at higher kinetic energy and at lower annealing temperatures~\cite{Onsten2009, Onsten2013, Grad2020}.

\begin{table}
	\caption{Interatomic distances on the reconstructed surface with neutral (V$_\mathrm{O}^{\bullet\bullet}$), singly (V$_\mathrm{O}^{\bullet}$) or doubly (V$_\mathrm{O}^\mathrm{X}$) positively  charged V$_\mathrm{O}$, compared to the stoichiometric surface and bulk \ce{Cu2O}. See Fig.~\ref{fig:stoichsurface} for atomic labels.}
	\begin{tabular*}{\columnwidth}{@{\extracolsep{\fill}}lccccc}
		\hline
		\hline
		Distance & V$_\mathrm{O}^{\bullet\bullet}$ (\AA) & V$_\mathrm{O}^{\bullet}$ (\AA) & V$_\mathrm{O}^\mathrm{X}$ (\AA) & Ideal (\AA) & Bulk (\AA) \\
		\hline
		Cu$_\mathrm{NN}$--Cu$_\mathrm{NN}$ & 2.47 & 2.47 & 2.47 & 3.01 & 3.01\\
		Cu$_\mathrm{NN}$--O$_\mathrm{CSA}$ & 1.93 & 1.94 & 1.94 & 1.86 & 1.85\\
		Cu$_\mathrm{CSA}$--O$_\mathrm{CUS}$ & - & - & - &1.81 & 1.85\\
		Cu$_\mathrm{CUS}$--O$_\mathrm{CSA}$ & 1.92 & 1.92 & 1.92 & 1.90 & 1.85\\
		Cu$_\mathrm{CUS}$-Cu$_\mathrm{NN}$ & 2.54& 2.64 & 2.58 & 3.01 &3.01\\
		\hline
	\end{tabular*}
	\label{tbl:surfrelaxations}
\end {table}
The relaxed structure of the stoichiometric surface has Cu$_\mathrm{CUS}$--O and Cu$_\mathrm{CSA}$--O$_\mathrm{CUS}$ bond lengths of about 1.90~\AA\ and 1.81~\AA, which are longer and shorter, respectively, compared to Cu--O bonds in bulk \ce{Cu2O} (see Table~\ref{tbl:surfrelaxations}). The formation of a neutral oxygen vacancy (V$_\mathrm{O}^{\bullet\bullet}$ in Kr\"oger-Vink notation~\cite{KROGER1956307}) due to removal of one O$_\mathrm{CUS}$ from the stoichiometric ($\sqrt{3} \times \sqrt{3}$)R30$^{\circ}$ surface results in a vacant site surrounded by three singly coordinated Cu ions (Cu$_\mathrm{NN}$, see Fig. \ref{fig:stoichsurface}c) and leads to large structural relaxations: Cu$_\mathrm{NN}$ move towards the V$_\mathrm{O}^{\bullet\bullet}$, forming a cluster of Cu ions with Cu$_\mathrm{NN}$--Cu$_\mathrm{NN}$ distances of about 2.47~\AA\ and at the same time, the Cu$_\mathrm{CUS}$ ions also move towards the defect, resulting in Cu$_\mathrm{CUS}$--Cu$_\mathrm{NN}$ distances of about 2.54~\AA, in both cases much shorter than the Cu--Cu bond lengths in the bulk (3.01~\AA, see Table~\ref{tbl:surfrelaxations}). This picture is in disagreement with semi-local DFT results that report V$_\mathrm{O}^{\bullet\bullet}$ formation not to cause appreciable distortions in the surface structure~\cite{SUN2008, Zhang2018}, but it is in line with DFT+$U$ results~\cite{Bendavid2013JPCB}. The formation of charged V$_\mathrm{O}$ was, however, not taken into account in these studies. We find that the singly positively  (V$_\mathrm{O}^{\bullet}$) and doubly positively (V$_\mathrm{O}^\mathrm{X}$)  charged vacancies result in structural relaxations similar to V$_\mathrm{O}^{\bullet\bullet}$, even though the distance between Cu$_\mathrm{CUS}$ and Cu$_\mathrm{NN}$ is slightly larger compared to V$_\mathrm{O}^{\bullet\bullet}$ (see Table~\ref{tbl:surfrelaxations} and Fig.~\ref{fig:stoichsurface}c).

\begin{figure}
 \centering
 \includegraphics[width=\columnwidth]{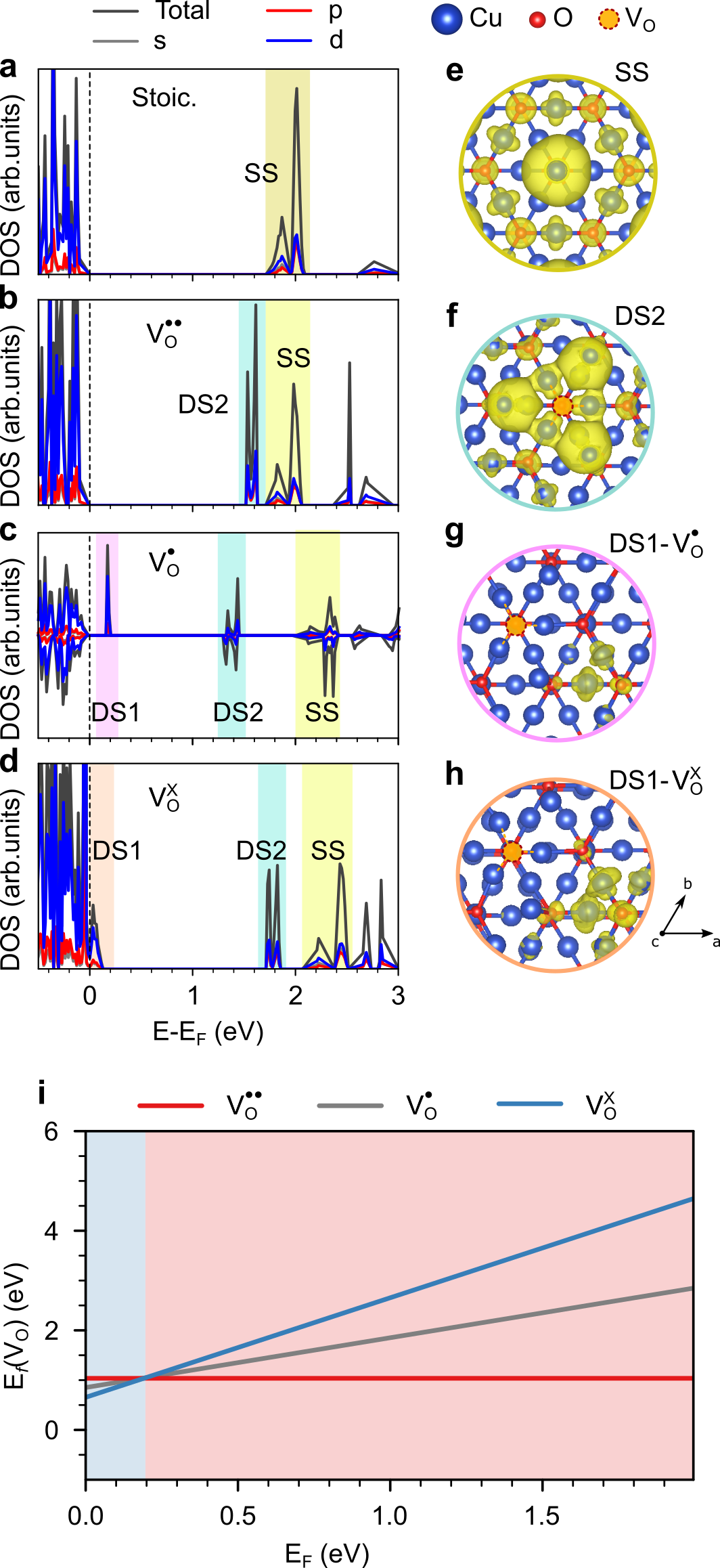}
 \caption{Density of states for the ($\sqrt{3} \times \sqrt{3}$)R30$^{\circ}$ supercell a) in the stoichiometric case and with one b) V$_\mathrm{O}^{\bullet\bullet}$, c) V$_\mathrm{O}^{\bullet}$, and d) V$_\mathrm{O}^\mathrm{X}$. The zero of the energy scale was set at the Fermi energy. For the spin-polarized V$_\mathrm{O}^{\bullet}$ calculation, the DOS for the spin-up and spin-down channels are reported with positive and negative values respectively on the $y$-axis. The isosurfaces (2$\times 10 ^{-2}$ e/\AA$^3$) in e-i) correspond to the charge density associated with the defect states (DS) or surface states (SS) highlighted with the corresponding color in plots a-d). l) Oxygen vacancy formation energy ($E_f(\mathrm{V_O})$) under O-poor conditions in different charge states as a function of the Fermi energy ranging from the valence band maximum ($E_F=0$) up to the experimental band gap for the \ce{Cu2O}-(111) surface.}
\label{fig:PDOS_surf}
\end{figure}
Figure~\ref{fig:PDOS_surf}a illustrates the electronic properties of the stoichiometric non-reconstructed \ce{Cu2O} (111) surface: in agreement with scanning tunneling spectroscopy (STS)~\cite{Onsten2009, Zhang2018} and photoemission~\cite{Grad2020} the band gap is significantly smaller than in bulk \ce{Cu2O}. This is due to the appearance of a peak at about 1.78~eV from the valence band maximum (VBM) that can be attributed to surface states localized mainly on Cu$_\mathrm{CUS}$ and Cu$_\mathrm{CSA}$ atoms in the outermost O--Cu--O trilayer (see Fig.~\ref{fig:PDOS_surf}e). 

On the oxygen deficient reconstructed surface, a V$_\mathrm{O}^{\bullet\bullet}$ results in the two excess electrons occupying states resonant with the valence band (VB, see Fig.~\ref{fig:PDOS_surf}b) and in the appearance of an empty localized defect state (DS2 in Fig.~\ref{fig:PDOS_surf}b) below the surface states (about 1.61~eV from VBM). DS2 has contributions of the three Cu$_\mathrm{NN}$ and the three Cu$_\mathrm{CUS}$ atoms closest to the oxygen vacancy (see Fig.~\ref{fig:PDOS_surf}f). If one oxygen atom and one electron are simultaneously removed to form V$_\mathrm{O}^{\bullet}$, the extra electron occupies states resonant with the VB, while an unoccupied defect state appears at 0.17~eV, mainly localized on two Cu atoms far from the V$_\mathrm{O}^{\bullet}$ (DS1 in Fig.~\ref{fig:PDOS_surf}c and g), similar to what we observe for bulk \ce{Cu2O} (SI Fig.~\ref{fig:PDOS_bulk_VO}). Furthermore, in presence of V$_\mathrm{O}^{\bullet}$, DS2 is stabilized and appears at about 1.3-1.4~eV in the gap, while the surface states are pushed up beyond 2.0~eV just below the conduction band (CB). Finally, V$_\mathrm{O}^\mathrm{X}$ is associated with the empty defect state DS1 merged with the top of the VB and mainly localized on Cu atoms far away from the defect (see Fig.~\ref{fig:PDOS_surf}d and h). The splitting between the DS2 and surface states is still visible, but the corresponding peaks appear at about 1.8 and 2.1~eV, respectively, the highest ones being merged with the CB.

Figure~\ref{fig:PDOS_surf}i shows the formation energies of these surface V$_\mathrm{O}$ in the different charge states as a function of the position of the Fermi energy under O-poor conditions. The neutral defect has a formation energy of about 1.2~eV, slightly lower than previously reported using standard DFT functionals (1.4-1.7~eV)~\cite{SUN2008, Bendavid2013JPCB, Zhang2018}. As in the bulk (see SI Section \ref{SISec:defbulk}), the neutral defect is most stable, except for Fermi energies just above the VBM, where V$_\mathrm{O}^\mathrm{X}$ becomes the favored charge state with a thermodynamic transition level $\epsilon(+2/+1)$ of 0.20~eV. This confirms the existence of doubly charged V$_\mathrm{O}^\mathrm{X}$ under \textit{p}-type doping, leading to the ($\sqrt{3} \times \sqrt{3}$)-R30$^{\circ}$ reconstruction. These vacancies strongly repel each other, simple electrostatic point-charge calculations suggesting energies larger by 4.5~eV for arrangements other than the ordered reconstruction.

\subsection{Carrier Capture}

Bulk calculations performed with better converged parameters compared to previous HSE calculations show that a thermodynamic transition level $\epsilon(0/+1)$ exists close to the valence band edge (see SI Section \ref{sec:defbulk}). Bulk oxygen vacancies could therefore, in principle, trap holes, while electron trapping is not possible. The computed hole-capture coefficient (see SI Section \ref{sec:capture}) is, however, very small, demonstrating that neither electron nor hole trapping by bulk V$_\mathrm{O}$ is likely.

\begin{figure}
	\centering
	\includegraphics[width=\columnwidth]{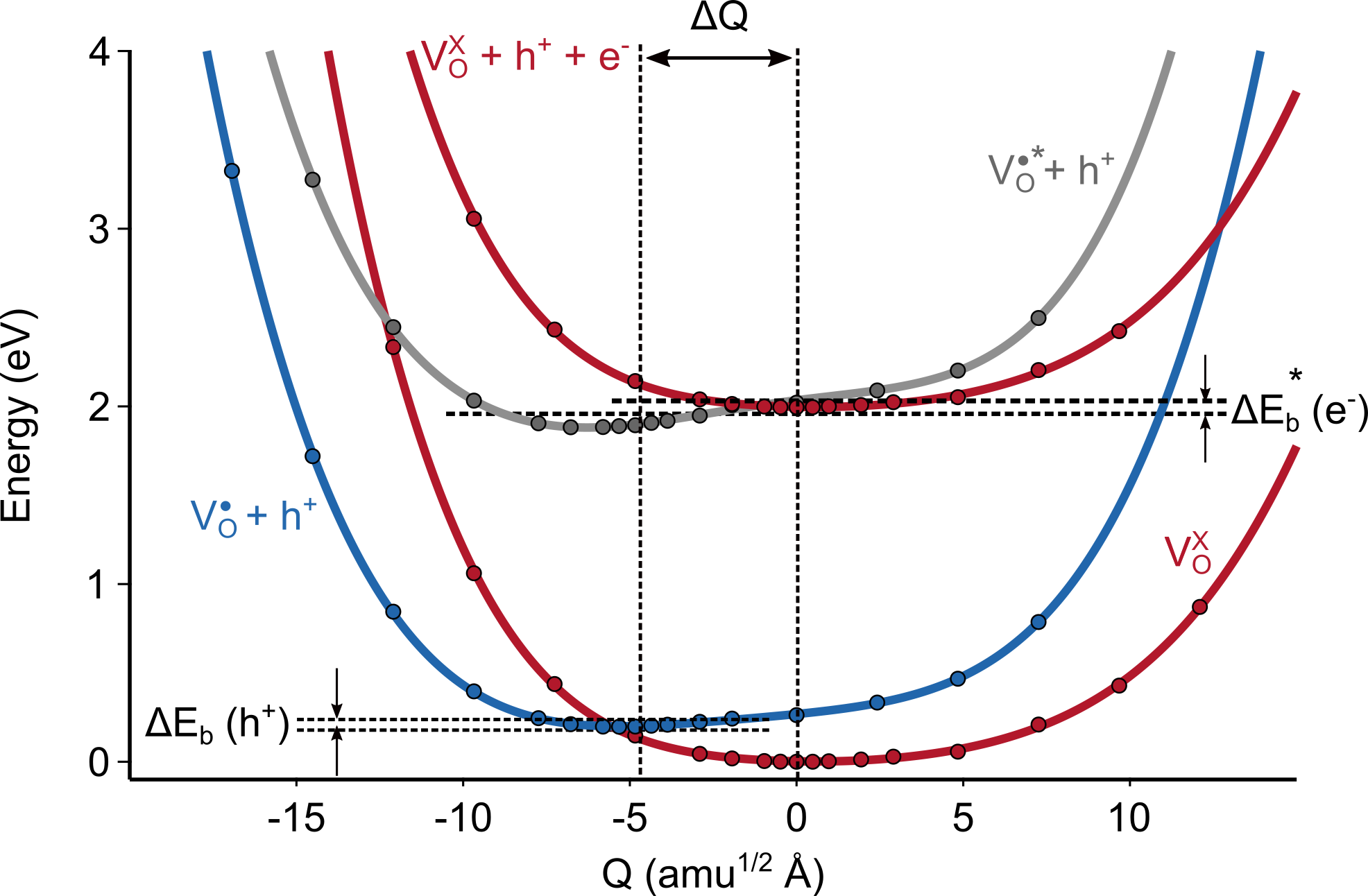}
	\caption{Configuration coordinate diagram for V$_\mathrm{O}$(+2/+1) carrier capture at the reconstructed \ce{Cu2O}-(111) surface. The solid circles represent the relative formation energies calculated using hybrid DFT and lines are spline fits. $\Delta$E$_\mathrm{b}$ are carrier capture barriers.}
	\label{fig:surfcarriercapture}
\end{figure}

The calculated 1D configuration coordinate diagram for carrier capture at the reconstructed ($\sqrt{3} \times \sqrt{3}$)-R30$^{\circ}$-(111) surface (Fig.~\ref{fig:surfcarriercapture}) shows a very flat potential energy landscape for both the ground and excited states. The ground state is the doubly charged V$_\mathrm{O}^\mathrm{X}$, while for the excited state we initially consider the singly charged V$_\mathrm{O}^{\bullet}+h^+$ with a hole in the VB. The minima of this ground and excited state are horizontally offset by $\Delta Q = 4.83~\mathrm{amu^{\sfrac{1}{2}}\AA}$ (see Table~\ref{tbl:surfcarriercapture}), due to the large relaxations involving in particular the Cu$_\mathrm{CUS}$ (see Table~\ref{tbl:surfrelaxations}), while the energy difference between the two minima is only 0.20~eV, corresponding to the $\epsilon(+2/+1)$ charge transition level. The flat landscape and the large lattice relaxations result in a small hole capture barrier since the two curves intersects only 0.003~eV above the minimum of V$_\mathrm{O}^{\bullet}+h^+$. This results in a hole-capture coefficient $C = 1.24\times10^{-19}~\mathrm{cm^3/s}$ and hole-capture cross section $\sigma = 7.87\times10^{-11}~\mathrm{\AA^2}$ at 298~K (see SI Fig.~\ref{fig:carriercapture_vs_T} for temperature-dependent values). While the potential energy curves of V$_\mathrm{O}^\mathrm{X}+h^+ + e^-$ and V$_\mathrm{O}^{\bullet}+h^+$ intersect, the electron-capture barrier in excess of 2 eV will, given the inverse exponential dependence, result in a very small non-radiative electron capture rate.
\begin{table}
	\scriptsize
	\caption{Key parameters for carrier capture processes at the reconstructed \ce{Cu2O} (111) surface: total mass-weighted distortions ($\Delta Q$, in $\mathrm{amu^{\sfrac{1}{2}}\AA}$), ionization energy ($\Delta E$, in eV), carrier capture barrier ($\Delta E_b$, in eV), carrier capture coefficient ($C$, in $\mathrm{cm^3/s}$) and carrier capture cross section ($\sigma$ in $\mathrm{\AA^2}$) at 298 K.}
	\begin{tabular*}{\columnwidth}{@{\extracolsep{\fill}}cccccccccc}
		\hline
		\hline
		Defect & Carrier & $\Delta Q$ &$\Delta E$ 	& $\Delta E_b$ & $C$ &$\sigma$ \\
		\hline
		(+2/+1) & h$^+$ & 4.83 & 0.20 & 0.003 & 1.24$\times$10$^{-19}$ & 7.87$\times$10$^{-11}$ \\
		(+2/+1*) & e$^-$ & 4.83 & 2.00 & 0.011 & 4.87$\times$10$^{-9}$ & 3.82 \\
		\hline
		\hline
	\end{tabular*}
	\label{tbl:surfcarriercapture}
\end {table}

\begin{figure}
 \centering
 \includegraphics[width=\columnwidth]{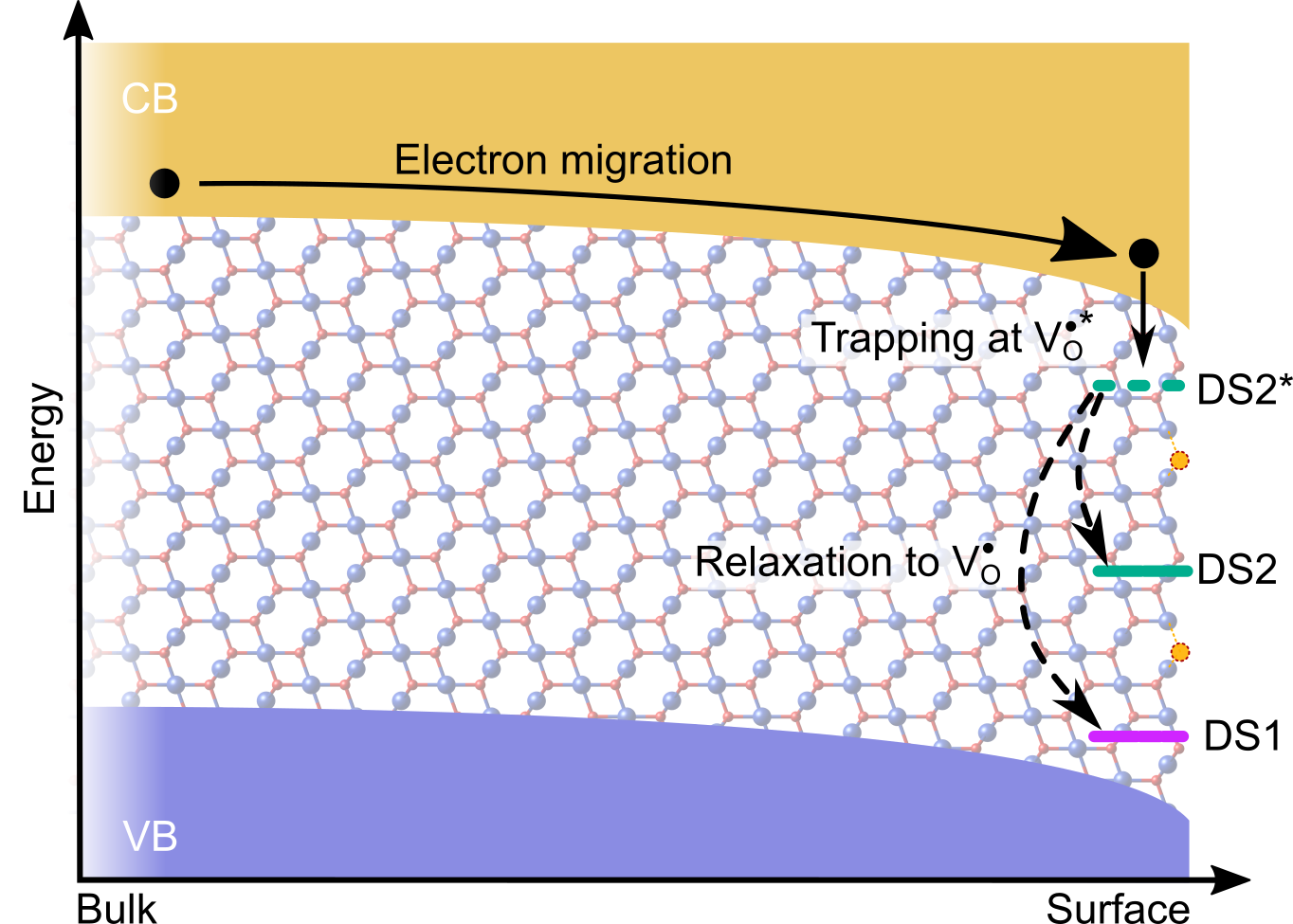}
 \caption{Schematic representation of electron capture at the reconstructed ($\sqrt{3} \times \sqrt{3}$)R30$^{\circ}$ \ce{Cu2O}-(111) surface. Due to surface band bending photo-excited electrons drift towards the surface, where they can either drive photocatalytic reactions (not shown) or be trapped in surface defect states (DS), if available. DS shown in the schematic result from experimentally observed surface oxygen vacancies (V$_{\mathrm{O}}$). Direct trapping into DS1 or DS2 is slow, but an electron can very rapidly be trapped into an excited state DS2*, from where it can relax into either DS1 or DS2. Blue and red spheres represents Cu and O atoms, respectively, while V$_{\mathrm{O}}$ are shown in orange.}
\label{fig:scheme}
\end{figure}

Excited states of defects are believed to play a pivotal role in multiphonon emission processes in wide-gap materials~\cite{AlkauskasPRB2016}. For a V$_\mathrm{O}^{\bullet}$ on the reconstructed surface, a spin-conserving excited state is reached by excitation of the extra electron from defect state DS1 to DS2 of the singly charged vacancy (see Fig.~\ref{fig:PDOS_surf}c). Beyond-DFT methods would, in principle, be necessary to describe excited states, but Alkauskas \textit{et al.}~\cite{AlkauskasPRB2016} showed that, for hybrid DFT functionals, accurate results can be obtained by approximating total energy differences by differences in single-particle Kohn-Sham eigenvalues in the spirit of the generalized Koopmans' theorem: at fixed geometry, the total energy of the excited state for a singly charged defect (V$_\mathrm{O}^{\bullet *}$) is higher than that of the ground state V$_\mathrm{O}^{\bullet}$ by the eigenvalue difference between DS2 and DS1 evaluated for the ionized V$_\mathrm{O}^\mathrm{X}$ state
\begin{equation}
E_\mathrm{tot, V_O^{\bullet *}} = E_\mathrm{tot, V_O^{\bullet}} + \epsilon_\mathrm{DS2, V_O^X} - \epsilon_\mathrm{DS1, V_O^X},
\label{eq:formenerg_exc}
\end{equation}
where all terms are consistently calculated for the same geometry. As shown in Fig. \ref{fig:surfcarriercapture}, when this excited state is taken into account, electron capture into V$_{\mathrm{O}}^{\bullet *}$ can occur because the (+2/+1*) transition lies closer to the CBM by the intradefect excitation energy of about 2.0~eV. The resulting electron-capture barrier of 0.011~eV and the proximity of the excited state to the CB lead to a large capture coefficient $C = 4.87\times10^{-9}~\mathrm{cm^3/s}$ and cross section $\sigma = 3.87~\mathrm{\AA^2}$ at 298~K (see SI Fig.~\ref{fig:carriercapture_vs_T} for temperature-dependent values). Considering a defect density of $4.58\times10^{21}~\mathrm{cm^{-3}}$ (one defect per supercell within 1~nm from the surface), this electron-capture coefficient results in an electron lifetime in the CB of about $4.47\times10^{-14}~\mathrm{s}$ and a mean free path of $5.71\times10^{-07}~\mathrm{cm}$. After the capture process the system can quickly relax from V$_{\mathrm{O}}^{\bullet *}$ to V$_{\mathrm{O}}^{\bullet}$ via intradefect relaxations, the extra electron occupying either DS1 or DS2. This electron-capture process is represented in Fig.~\ref{fig:scheme}.
\begin{figure}
 \centering
 \includegraphics[width=\columnwidth]{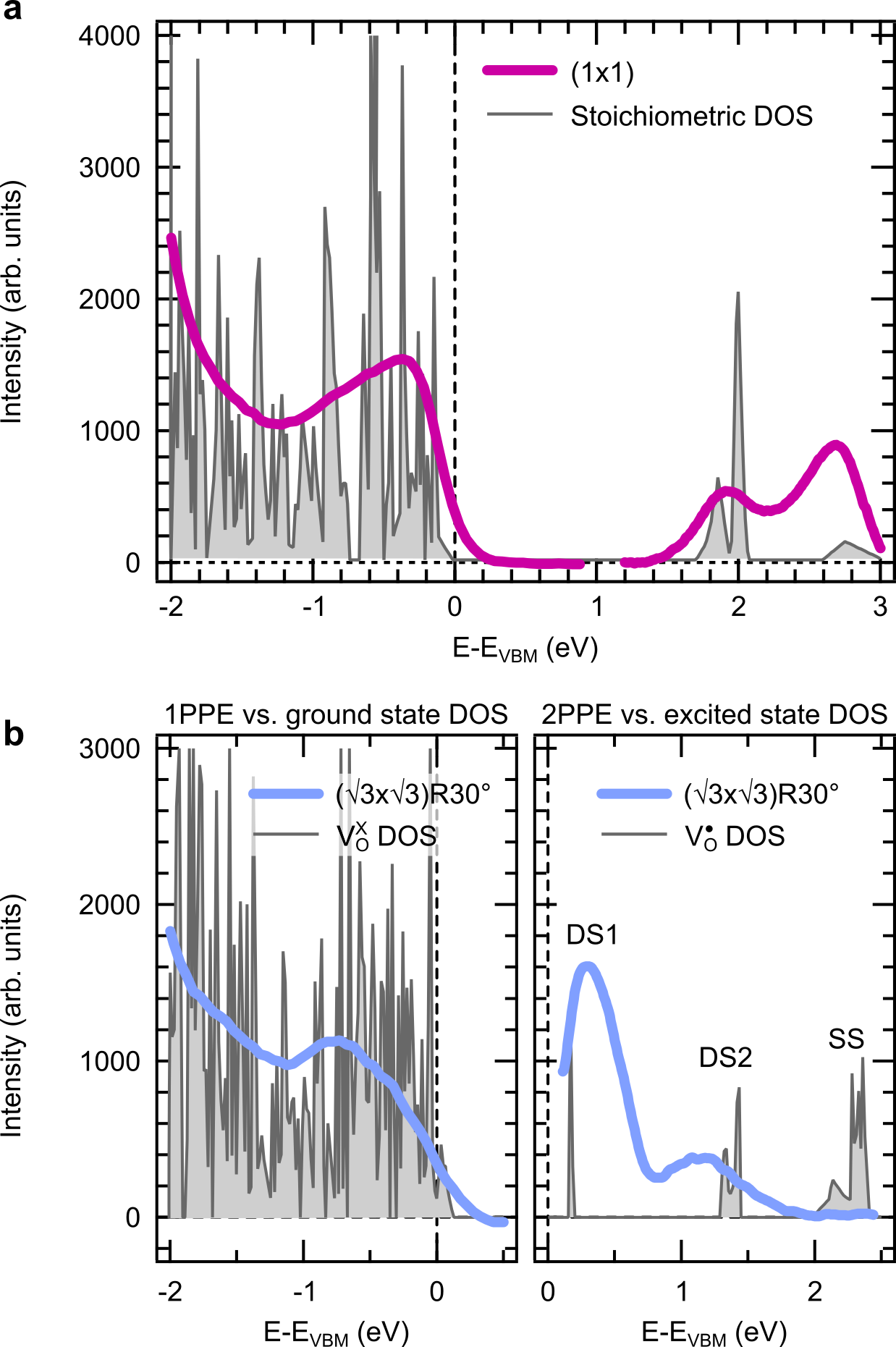}
 \caption{Comparison of the experimentally measured (dashed lines) and computed (solid lines) density of states for a) the stoichiometric ($1 \times 1$) and b) the reconstructed ($\sqrt{3} \times \sqrt{3}$)-R30$^{\circ}$-(111) \ce{Cu2O} surface with V$_{\mathrm{O}}$. For the valence band, determined using 1PPE, we compare with the ground-state V$_\mathrm{O}^\mathrm{X}$ DOS, whereas for the empty states, determined using 2PPE, we compare with the excited-state V$_\mathrm{O}^{\bullet}$ DOS. For the spin-polarized V$_\mathrm{O}^{\bullet}$ DOS, the spin-up and spin-down channels have been summed. Note the missing photoelectron signals at energies higher than 2~eV in (b).}
\label{fig:exptdos}
\end{figure}
This mechanism for electron capture can be confirmed comparing the theoretical results with the electron dynamics at the (111) surface of \ce{Cu2O} experimentally determined via time-resolved 2PPE~\cite{Grad2020}. In these experiments, electrons are excited from the VB to the CB by absorption of a 3~eV laser pulse. The energy distribution of photoexcited electrons above the Fermi energy is monitored by a 6~eV laser pulse as function of the time delay between both pulses. In addition, valence band spectra were measured using ultraviolet photoelectron spectroscopy with He I$_\alpha$ radiation (21.2~eV). In Fig.~\ref{fig:exptdos} the measured valence and conduction band spectra are shown together with the computed DOS for both surfaces. On the stoichiometric ($1 \times 1$) surface, the surface and low-lying conduction band states at 1.9~eV and 2.6~eV above the VBM are present, in good agreement with the computed DOS. The lifetime of excited electrons in the CB was determined as 10~ps. However, for the reconstructed ($\sqrt{3} \times \sqrt{3}$)-R30$^{\circ}$ surface no photoemission signal from the CB is observed, but an intense defect state at 0.1~eV and another weaker state at 0.9~eV dominate the spectra. The missing photoemission signal from the CB is evidence for an ultrafast capture of photoexcited electrons by defects on the reconstructed surface. Occupation of those defect states by the excited electrons is faster than filling the low-lying CB states (0.030~ps) in good agreement with the electron lifetime in the CB predicted by DFT.  These defects act as traps, since no noticeable decay of their electron population is observed over a time scale of hundreds of picoseconds. The good agreement between the experimental spectrum obtained in the 2PPE by populating the ground state V$_\mathrm{O}^\mathrm{X}$ with one extra electron and the computed density of states for the reconstructed surface with V$_{\mathrm{O}}^{\bullet}$ with in Fig.~\ref{fig:exptdos} confirms the predominance of charged oxygen vacancies in the ($\sqrt{3} \times \sqrt{3}$)-R30$^{\circ}$ reconstruction of the \ce{Cu2O} (111) surface, as well as the crucial role of the  V$_{\mathrm{O}}^\mathrm{X}$/V$_{\mathrm{O}}^{\bullet *}$ transition for the observed efficient electron-capture at this surfaces.

\section{Conclusions}

In summary, our hybrid DFT results show that bulk defects in \ce{Cu2O} cannot trap electrons and are highly inefficient hole traps. We further confirm that the $\sqrt{3} \times \sqrt{3}$)-R30$^{\circ}$ reconstruction experimentally observed under oxygen-poor conditions on the \ce{Cu2O} (111) surface is associated with a $\sfrac{1}{3}$ monolayer of charged oxygen vacancies, as previously hypothesized based on experiments. We report an excellent agreement between experimental spectra and computed densities of states that confirm the singly charged (V$_{\mathrm{O}}^{\bullet}$) state of these defects after electron trapping, and hence support the computed doubly charged V$_{\mathrm{O}}^\mathrm{X}$ ground-state. Due to electrostatic repulsion, these defects form a highly ordered structure that strongly suppresses electron transport. In particular, the excited state of the V$_{\mathrm{O}}^{\bullet}$ plays a crucial role in efficient electron capture at the surface. Electrons are trapped by a transition from V$_{\mathrm{O}}^\mathrm{X}$ to the excited state V$_{\mathrm{O}}^{\bullet *}$. After the capture process, the system can relax to V$_{\mathrm{O}}^{\bullet }$ via intradefect relaxation, the electron finally being trapped in the defect states associated with the singly charged defect. This process is predicted to have a capture coefficient of about $10^{-9}~\mathrm{cm^3/s}$ leading to a free-carrier lifetime of 0.04~ps, in good agreement with experimental observations. The findings demonstrate the predominance of surface oxygen vacancies in limiting the photocurrent of \ce{Cu2O} based heterostructures and provide a solid basis for the engineering of more efficient \ce{Cu2O} photoelectrodes.

\section*{Acknowledgments}
This work was funded by the Swiss National Science Foundation (grant numbers 200020\_172641, PP00P2\_157615 and PP00P2\_187185). Computational resources for this work were provided by the University of Bern (on the HPC cluster UBELIX, \url{http://www.id.unibe.ch/hpc}) and by the Swiss National Supercomputing Center (CSCS) under project IDs s955 and s1033. 

\section*{Author contributions}
C.R. performed DFT calculations, L.G. performed experiments. C.R. and U.A. wrote the manuscript. All authors jointly analyzed the data and revised the manuscript. 

\section*{Competing interests}
The authors declare no competing interests.

\section{Methods}\label{sec:compdetails}

All density functional theory (DFT) calculations were performed with the Vienna Ab-initio Simulation Package (VASP) ~\cite{Kresse199347, Kresse1994, KRESSE199615, Kresse199654}. The Heyd-Scuseria-Ernzerhof (HSE)~\cite{HeydJCP2003, HeydJCP2004} hybrid exchange-correlation functional was employed together with the default VASP Cu and O projector augmented wave (PAW) potentials~\cite{BlochlPRB1994,KressePRB1999} with Cu(4s, 3d) and O(2s, 2p) valence configurations. Wavefunctions were expanded in planewaves up to a kinetic energy of 500 eV. This setup provides an excellent agreement with experiment (see SI Sections~\ref{sec:compbulk} and \ref{SISec:defbulk}).

Surface calculations were performed using a ($\sqrt{3} \times \sqrt{3}$)R30$^{\circ}$ supercell of the ideal O-terminated (111) surface from which one surface O atom was removed, corresponding to a \sfrac{1}{3} V$_\mathrm{O}$ monolayer. The slab model consists of five O-Cu-O trilayers with a 10~\AA\ vacuum gap to prevent interactions between periodic images along the surface normal. The Brillouin zone of this slab model was sampled using a $3\times3\times1$ \textbf{k}-point mesh. Different charge states for the V$_\mathrm{O}$ were studied by adjusting the number of electrons and by applying a neutralizing background charge, as required by calculations under periodic-boundary conditions to avoid a divergence of the electrostatic potential. Atomic positions were optimized while keeping the lattice vectors fixed at optimized values of the non-defective system. Additionally, the atoms in the bottom two trilayers were kept fixed at bulk positions. Structural relaxations were performed until forces converged below 10$^{-3}$~eV/\AA.

The formation energy of a V$_\mathrm{O}$ in a charge state $q$ ($E_{\mathrm{f,V_O}^q}$) was calculated as described in Ref.~\onlinecite{freysoldt2014first}:
\begin{multline}
E_{\mathrm{f,V_O}^q} = E_{\mathrm{tot,V_O}^q} - E_{\mathrm{tot,stoich}} + n_\mathrm{O} \mu_\mathrm{O} \\
+ q[E_{\mathrm{VBM}} + E_\mathrm{F}] + E_{\mathrm{corr}}\,
\label{eq:formenerg}
\end{multline}
where $E_{\mathrm{tot,V_O}^q}$ and $E\mathrm{_{tot,stoich}}$ are the total energies of the defective and stoichiometric supercells, respectively, and $E\mathrm{_F}$ is the Fermi energy relative to the valence band maximum ($E\mathrm{_{VBM}}$) of the stoichiometric cell, which can assume values within the band gap $E\mathrm{_g}$ ($0 \leq E\mathrm{_F} \leq E\mathrm{_g}$) of the non-defective structure. The third term includes the number of removed O atoms ($n_\mathrm{O} < 0$) and the oxygen chemical potential $\mu_\mathrm{O}$. Results are reported for oxygen-poor conditions: $\mu_{\textrm{O}}=\frac{1}{2}\mu_{\textrm{O$_2$}} + \Delta \mu_{\textrm{O}}$, where $\mu_{\textrm{O$_2$}}$ is the energy of the \ce{O2} molecule and $\Delta \mu_{\textrm{O}}= -1.55$ eV (see SI Section \ref{sec:compchempot} for details). Lastly, $E_{\textrm{corr}}$ is a corrective term necessary to align the electrostatic potentials of the charged defective and the neutral stoichiometric cell obtained by averaging the electrostatic potential in spheres around atomic sites located far from the defect~\cite{lany2008}. No further finite-size corrections were applied since the defect concentration we simulate is realistic for this surface. 

The thermodynamic transition level $\epsilon (q_1/q_2)$ for V$_\mathrm{O}$ defects with charge states $q_1$ and $q_2$ was computed as the Fermi level for which the formation energies of the two charge states are equal~\cite{freysoldt2014first}:
\begin{equation}
\epsilon(q_1/q_2) = \frac{E_{\mathrm{f,V_O}^{q_1}}(E_\mathrm{F}=0) - E_{\mathrm{f,V_O}^{q_2}}(E_\mathrm{F}=0) }{q_2 -q_1}\,
\label{eq:transitionlevel}
\end{equation}
where $E_{\mathrm{f,V_O}^{q}}$ is the formation energy of a V$_\mathrm{O}$ in a charge state $q$ in its equilibrium structure when the Fermi level is at the valence-band maximum ($E\mathrm{_F=0}$). For Fermi-level positions below $\epsilon (q_1/q_2)$, charge state $q_1$ is stable, while for Fermi-level positions above $\epsilon (q_1/q_2)$, charge state $q_2$ is stable.

We computed surface energies via \textit{ab initio} atomistic thermodynamics, as shown for \ce{Cu2O} surfaces in Ref.~\onlinecite{Bendavid2013JPCB}:
\begin{equation}
\gamma=\frac{1}{2A}\left[E_{\mathrm{slab}}-N_\mathrm{Cu} \mu_{\mathrm{Cu}} - N_\mathrm{O} \mu_{\mathrm{O}}\right],
\label{eq:nonstoichsurfen}
\end{equation}
where $E_{\mathrm{slab}}$ is the total energy of the slab, $N_\mathrm{Cu}$/$N_\mathrm{O}$ are the number of Cu/O atoms in the model and $\mu_\mathrm{Cu}$/$\mu_\mathrm{O}$ are the Cu/O chemical potentials. $A$ is the surface area and the $1/2$ factor accounts for the two surfaces contained in the slab. Result are reported for O-rich ($\Delta \mu_{\textrm{O}}= -1.41$~eV and $\Delta \mu_{\textrm{Cu}}= -0.07$~eV) and O-poor conditions ($\Delta \mu_\textrm{Cu} =0$ and $\Delta \mu_\textrm{O} =-1.55$~eV, see SI Sections~\ref{sec:compchempot} for details. 

We considered non-radiative carrier capture processes involving a defect and occurring via multi-phonon emission~\cite{Huang1950, HenryLang1977}. Such processes can be described in DFT using the approach introduced by Alkauskas \textit{et al.}~\cite{AlkauskasPRB2014, AlkauskasPRB2016} based on the static approximation~\cite{Huang1981} and using an effective one-dimensional configuration coordinate $Q$ to represent the phonon wavefunction of all vibrations coupling to the change of the defect geometry upon carrier capture. Within this approach, the capture coefficient is
\begin{multline}
C(T)=sV \eta_{sp} g \frac{2\pi }{\hbar}W_{if}^2 \sum_{m,n} w_m(T) \vert\langle \chi_{im} \vert Q + \Delta Q \vert \chi_{fn} \rangle \vert ^2 \\
\times \delta( \Delta E + \epsilon_{im} -\epsilon_{fn}) ,
\label{eq:capturecoeff}
\end{multline}
where $V$ is the volume of the supercell, $\eta_{sp} $ accounts for spin-selection rules ($\eta_{sp}=1/2 $ when the initial state is a spin doublet and the final state is a spin singlet), $g$ is the degeneracy of the final state, $W_{if}$ is the electron-phonon coupling matrix element of the initial and final states, $\Delta E$ is the energy difference between the two states, $\chi$ and $\epsilon$ are the phonon wave functions and eigenvalues, respectively, for the excited ($im$) and ground ($fn$) electronic states, while $w_m$ is the thermal occupation number of the excited vibrational state. $\delta$ is replaced with a Gaussian function of finite width. Matrix elements $W_{if}$ are computed according to Ref.~\cite{AlkauskasPRB2014} using the \texttt{PAWpySeed} package~\cite{pawpyseed} to obtain the wavefunction overlap. The Sommerfeld factor $s$ was computed according to Ref.~\cite{KimJmatChemA2019, KimRSC2020}, requiring electron and hole effective masses, obtained by finite differences as implemented in EMC~\cite{fonari2012effective}. Non-radiative capture coefficients were calculated with the \texttt{CarrierCapture.jl} package~\cite{carriercapture}. Additional information can be found in SI Section \ref{sec:compcapture}.

\section*{Data availability}
Data associated with the calculations is available on the Materials Cloud: \url{https://doi.org/10.24435/materialscloud:rr-2n}.

\bibliography{references}

\begin{thebibliography}{41}%
\makeatletter
\providecommand \@ifxundefined [1]{%
 \@ifx{#1\undefined}
}%
\providecommand \@ifnum [1]{%
 \ifnum #1\expandafter \@firstoftwo
 \else \expandafter \@secondoftwo
 \fi
}%
\providecommand \@ifx [1]{%
 \ifx #1\expandafter \@firstoftwo
 \else \expandafter \@secondoftwo
 \fi
}%
\providecommand \natexlab [1]{#1}%
\providecommand \enquote  [1]{``#1''}%
\providecommand \bibnamefont  [1]{#1}%
\providecommand \bibfnamefont [1]{#1}%
\providecommand \citenamefont [1]{#1}%
\providecommand \href@noop [0]{\@secondoftwo}%
\providecommand \href [0]{\begingroup \@sanitize@url \@href}%
\providecommand \@href[1]{\@@startlink{#1}\@@href}%
\providecommand \@@href[1]{\endgroup#1\@@endlink}%
\providecommand \@sanitize@url [0]{\catcode `\\12\catcode `\$12\catcode
  `\&12\catcode `\#12\catcode `\^12\catcode `\_12\catcode `\%12\relax}%
\providecommand \@@startlink[1]{}%
\providecommand \@@endlink[0]{}%
\providecommand \url  [0]{\begingroup\@sanitize@url \@url }%
\providecommand \@url [1]{\endgroup\@href {#1}{\urlprefix }}%
\providecommand \urlprefix  [0]{URL }%
\providecommand \Eprint [0]{\href }%
\providecommand \doibase [0]{http://dx.doi.org/}%
\providecommand \selectlanguage [0]{\@gobble}%
\providecommand \bibinfo  [0]{\@secondoftwo}%
\providecommand \bibfield  [0]{\@secondoftwo}%
\providecommand \translation [1]{[#1]}%
\providecommand \BibitemOpen [0]{}%
\providecommand \bibitemStop [0]{}%
\providecommand \bibitemNoStop [0]{.\EOS\space}%
\providecommand \EOS [0]{\spacefactor3000\relax}%
\providecommand \BibitemShut  [1]{\csname bibitem#1\endcsname}%
\let\auto@bib@innerbib\@empty
\bibitem [{\citenamefont {Meyer}\ \emph {et~al.}(2012)\citenamefont {Meyer},
  \citenamefont {Polity}, \citenamefont {Reppin}, \citenamefont {Becker},
  \citenamefont {Hering}, \citenamefont {Klar}, \citenamefont {Sander},
  \citenamefont {Reindl}, \citenamefont {Benz}, \citenamefont {Eickhoff},
  \citenamefont {Heiliger}, \citenamefont {Heinemann}, \citenamefont
  {Bl{\"a}sing}, \citenamefont {Krost}, \citenamefont {Shokovets},
  \citenamefont {M{\"u}ller},\ and\ \citenamefont {Ronning}}]{Meyer2012}%
  \BibitemOpen
  \bibfield  {author} {\bibinfo {author} {\bibfnamefont {B.~K.}\ \bibnamefont
  {Meyer}}, \bibinfo {author} {\bibfnamefont {A.}~\bibnamefont {Polity}},
  \bibinfo {author} {\bibfnamefont {D.}~\bibnamefont {Reppin}}, \bibinfo
  {author} {\bibfnamefont {M.}~\bibnamefont {Becker}}, \bibinfo {author}
  {\bibfnamefont {P.}~\bibnamefont {Hering}}, \bibinfo {author} {\bibfnamefont
  {P.~J.}\ \bibnamefont {Klar}}, \bibinfo {author} {\bibfnamefont {Th.}\
  \bibnamefont {Sander}}, \bibinfo {author} {\bibfnamefont {C.}~\bibnamefont
  {Reindl}}, \bibinfo {author} {\bibfnamefont {J.}~\bibnamefont {Benz}},
  \bibinfo {author} {\bibfnamefont {M.}~\bibnamefont {Eickhoff}}, \bibinfo
  {author} {\bibfnamefont {C.}~\bibnamefont {Heiliger}}, \bibinfo {author}
  {\bibfnamefont {M.}~\bibnamefont {Heinemann}}, \bibinfo {author}
  {\bibfnamefont {J.}~\bibnamefont {Bl{\"a}sing}}, \bibinfo {author}
  {\bibfnamefont {A.}~\bibnamefont {Krost}}, \bibinfo {author} {\bibfnamefont
  {S.}~\bibnamefont {Shokovets}}, \bibinfo {author} {\bibfnamefont
  {C.}~\bibnamefont {M{\"u}ller}}, \ and\ \bibinfo {author} {\bibfnamefont
  {C.}~\bibnamefont {Ronning}},\ }\bibfield  {title} {\enquote {\bibinfo
  {title} {{B}inary {C}opper {O}xide {S}emiconductors: {F}rom {M}aterials
  towards {D}evices},}\ }\href {\doibase 10.1002/pssb.201248128} {\bibfield
  {journal} {\bibinfo  {journal} {Phys. Status Solidi (B)}\ }\textbf {\bibinfo
  {volume} {249}},\ \bibinfo {pages} {1487--1509} (\bibinfo {year}
  {2012})}\BibitemShut {NoStop}%
\bibitem [{\citenamefont {Niu}\ \emph {et~al.}(2018)\citenamefont {Niu},
  \citenamefont {Moehl}, \citenamefont {Cui}, \citenamefont {Wick-Joliat},
  \citenamefont {Zhu},\ and\ \citenamefont {Tilley}}]{Niu2018}%
  \BibitemOpen
  \bibfield  {author} {\bibinfo {author} {\bibfnamefont {W.}~\bibnamefont
  {Niu}}, \bibinfo {author} {\bibfnamefont {T.}~\bibnamefont {Moehl}}, \bibinfo
  {author} {\bibfnamefont {W.}~\bibnamefont {Cui}}, \bibinfo {author}
  {\bibfnamefont {R.}~\bibnamefont {Wick-Joliat}}, \bibinfo {author}
  {\bibfnamefont {L.}~\bibnamefont {Zhu}}, \ and\ \bibinfo {author}
  {\bibfnamefont {S.~D.}\ \bibnamefont {Tilley}},\ }\bibfield  {title}
  {\enquote {\bibinfo {title} {{E}xtended {L}ight {H}arvesting with {D}ual
  \ce{Cu2O}-based {P}hotocathodes for {H}igh {E}fficiency {W}ater
  {S}plitting},}\ }\href {\doibase https://doi.org/10.1002/aenm.201702323}
  {\bibfield  {journal} {\bibinfo  {journal} {Adv. Energy Mater.}\ }\textbf
  {\bibinfo {volume} {8}},\ \bibinfo {pages} {1702323} (\bibinfo {year}
  {2018})}\BibitemShut {NoStop}%
\bibitem [{\citenamefont {Pan}\ \emph {et~al.}(2018)\citenamefont {Pan},
  \citenamefont {Kim}, \citenamefont {Mayer}, \citenamefont {Son},
  \citenamefont {Ummadisingu}, \citenamefont {Lee}, \citenamefont {Hagfeldt},
  \citenamefont {Luo},\ and\ \citenamefont {Gr{\"a}tzel}}]{Pan2018}%
  \BibitemOpen
  \bibfield  {author} {\bibinfo {author} {\bibfnamefont {L.}~\bibnamefont
  {Pan}}, \bibinfo {author} {\bibfnamefont {J.~H.}\ \bibnamefont {Kim}},
  \bibinfo {author} {\bibfnamefont {M.~T}\ \bibnamefont {Mayer}}, \bibinfo
  {author} {\bibfnamefont {M.-K.}\ \bibnamefont {Son}}, \bibinfo {author}
  {\bibfnamefont {A.}~\bibnamefont {Ummadisingu}}, \bibinfo {author}
  {\bibfnamefont {J.~S.}\ \bibnamefont {Lee}}, \bibinfo {author} {\bibfnamefont
  {A.}~\bibnamefont {Hagfeldt}}, \bibinfo {author} {\bibfnamefont
  {J.}~\bibnamefont {Luo}}, \ and\ \bibinfo {author} {\bibfnamefont
  {M.}~\bibnamefont {Gr{\"a}tzel}},\ }\bibfield  {title} {\enquote {\bibinfo
  {title} {{B}oosting the {P}erformance of \ce{Cu2O} {P}hotocathodes for
  {U}nassisted {S}olar {W}ater {S}plitting {D}evices},}\ }\href {\doibase
  10.1038/s41929-018-0077-6} {\bibfield  {journal} {\bibinfo  {journal} {Nat.
  Catal.}\ }\textbf {\bibinfo {volume} {1}},\ \bibinfo {pages} {412--420}
  (\bibinfo {year} {2018})}\BibitemShut {NoStop}%
\bibitem [{\citenamefont {Borgwardt}\ \emph {et~al.}(2019)\citenamefont
  {Borgwardt}, \citenamefont {Omelchenko}, \citenamefont {Favaro},
  \citenamefont {Plate}, \citenamefont {H{\"o}hn}, \citenamefont {Abou-Ras},
  \citenamefont {Schwarzburg}, \citenamefont {Van~de Krol}, \citenamefont
  {Atwater}, \citenamefont {Lewis}, \citenamefont {Eichberger},\ and\
  \citenamefont {Friedrich}}]{Borgwardt2019}%
  \BibitemOpen
  \bibfield  {author} {\bibinfo {author} {\bibfnamefont {M.}~\bibnamefont
  {Borgwardt}}, \bibinfo {author} {\bibfnamefont {S.}~\bibnamefont
  {Omelchenko}}, \bibinfo {author} {\bibfnamefont {M.}~\bibnamefont {Favaro}},
  \bibinfo {author} {\bibfnamefont {P.}~\bibnamefont {Plate}}, \bibinfo
  {author} {\bibfnamefont {C.}~\bibnamefont {H{\"o}hn}}, \bibinfo {author}
  {\bibfnamefont {D.}~\bibnamefont {Abou-Ras}}, \bibinfo {author}
  {\bibfnamefont {K.}~\bibnamefont {Schwarzburg}}, \bibinfo {author}
  {\bibfnamefont {R.}~\bibnamefont {Van~de Krol}}, \bibinfo {author}
  {\bibfnamefont {H.}~\bibnamefont {Atwater}}, \bibinfo {author} {\bibfnamefont
  {N.}~\bibnamefont {Lewis}}, \bibinfo {author} {\bibfnamefont
  {R.}~\bibnamefont {Eichberger}}, \ and\ \bibinfo {author} {\bibfnamefont
  {D.}~\bibnamefont {Friedrich}},\ }\bibfield  {title} {\enquote {\bibinfo
  {title} {{F}emtosecond {T}ime-{R}esolved {T}wo-{P}hoton {P}hotoemission
  {S}tudies of {U}ltrafast {C}arrier {R}elaxation in \ce{Cu2O}
  {P}hotoelectrodes},}\ }\href {\doibase 10.1038/s41467-019-10143-x} {\bibfield
   {journal} {\bibinfo  {journal} {Nat. Commun.}\ }\textbf {\bibinfo {volume}
  {10}},\ \bibinfo {pages} {1--7} (\bibinfo {year} {2019})}\BibitemShut
  {NoStop}%
\bibitem [{\citenamefont {Grad}\ \emph {et~al.}(2020)\citenamefont {Grad},
  \citenamefont {Novotny}, \citenamefont {Hengsberger},\ and\ \citenamefont
  {Osterwalder}}]{Grad2020}%
  \BibitemOpen
  \bibfield  {author} {\bibinfo {author} {\bibfnamefont {L.}~\bibnamefont
  {Grad}}, \bibinfo {author} {\bibfnamefont {Z.}~\bibnamefont {Novotny}},
  \bibinfo {author} {\bibfnamefont {M.}~\bibnamefont {Hengsberger}}, \ and\
  \bibinfo {author} {\bibfnamefont {J.}~\bibnamefont {Osterwalder}},\
  }\bibfield  {title} {\enquote {\bibinfo {title} {{I}nfluence of {S}urface
  {D}efect {D}ensity on the {U}ltrafast {H}ot {C}arrier {R}elaxation and
  {T}ransport in \ce{Cu2O} {P}hotoelectrodes},}\ }\href {\doibase
  10.1038/s41598-020-67589-z} {\bibfield  {journal} {\bibinfo  {journal} {Sci.
  Rep.}\ }\textbf {\bibinfo {volume} {10}},\ \bibinfo {pages} {1--10} (\bibinfo
  {year} {2020})}\BibitemShut {NoStop}%
\bibitem [{\citenamefont {Schulz}\ and\ \citenamefont
  {Cox}(1991)}]{SchulzCox1991}%
  \BibitemOpen
  \bibfield  {author} {\bibinfo {author} {\bibfnamefont {K.~H.}\ \bibnamefont
  {Schulz}}\ and\ \bibinfo {author} {\bibfnamefont {D.~F.}\ \bibnamefont
  {Cox}},\ }\bibfield  {title} {\enquote {\bibinfo {title} {{P}hotoemission and
  {L}ow-{E}nergy-{E}lectron-{D}iffraction {S}tudy of {C}lean and
  {O}xygen-{D}osed \ce{Cu2O} (111) and (100) {S}urfaces},}\ }\href {\doibase
  10.1103/PhysRevB.43.1610} {\bibfield  {journal} {\bibinfo  {journal} {Phys.
  Rev. B}\ }\textbf {\bibinfo {volume} {43}},\ \bibinfo {pages} {1610--1621}
  (\bibinfo {year} {1991})}\BibitemShut {NoStop}%
\bibitem [{\citenamefont {{\"O}nsten}\ \emph {et~al.}(2009)\citenamefont
  {{\"O}nsten}, \citenamefont {G{\"o}thelid},\ and\ \citenamefont
  {Karlsson}}]{Onsten2009}%
  \BibitemOpen
  \bibfield  {author} {\bibinfo {author} {\bibfnamefont {A.}~\bibnamefont
  {{\"O}nsten}}, \bibinfo {author} {\bibfnamefont {M.}~\bibnamefont
  {G{\"o}thelid}}, \ and\ \bibinfo {author} {\bibfnamefont {U.~O.}\
  \bibnamefont {Karlsson}},\ }\bibfield  {title} {\enquote {\bibinfo {title}
  {{A}tomic {S}tructure of \ce{Cu2O}(111)},}\ }\href {\doibase
  https://doi.org/10.1016/j.susc.2008.10.048} {\bibfield  {journal} {\bibinfo
  {journal} {Surf. Sci.}\ }\textbf {\bibinfo {volume} {603}},\ \bibinfo {pages}
  {257--264} (\bibinfo {year} {2009})}\BibitemShut {NoStop}%
\bibitem [{\citenamefont {{\"O}nsten}\ \emph {et~al.}(2013)\citenamefont
  {{\"O}nsten}, \citenamefont {Weissenrieder}, \citenamefont {Stoltz},
  \citenamefont {Yu}, \citenamefont {G{\"o}thelid},\ and\ \citenamefont
  {Karlsson}}]{Onsten2013}%
  \BibitemOpen
  \bibfield  {author} {\bibinfo {author} {\bibfnamefont {A.}~\bibnamefont
  {{\"O}nsten}}, \bibinfo {author} {\bibfnamefont {J.}~\bibnamefont
  {Weissenrieder}}, \bibinfo {author} {\bibfnamefont {D.}~\bibnamefont
  {Stoltz}}, \bibinfo {author} {\bibfnamefont {S.}~\bibnamefont {Yu}}, \bibinfo
  {author} {\bibfnamefont {M.}~\bibnamefont {G{\"o}thelid}}, \ and\ \bibinfo
  {author} {\bibfnamefont {U.~O.}\ \bibnamefont {Karlsson}},\ }\bibfield
  {title} {\enquote {\bibinfo {title} {{R}ole of {D}efects in {S}urface
  {C}hemistry on \ce{Cu2O}(111)},}\ }\href {\doibase 10.1021/jp3112217}
  {\bibfield  {journal} {\bibinfo  {journal} {J. Phys. Chem. C}\ }\textbf
  {\bibinfo {volume} {117}},\ \bibinfo {pages} {19357--19364} (\bibinfo {year}
  {2013})}\BibitemShut {NoStop}%
\bibitem [{\citenamefont {Sun}\ \emph {et~al.}(2008)\citenamefont {Sun},
  \citenamefont {Chen}, \citenamefont {Zheng},\ and\ \citenamefont
  {Lu}}]{SUN2008}%
  \BibitemOpen
  \bibfield  {author} {\bibinfo {author} {\bibfnamefont {B.-Z.}\ \bibnamefont
  {Sun}}, \bibinfo {author} {\bibfnamefont {W.-K.}\ \bibnamefont {Chen}},
  \bibinfo {author} {\bibfnamefont {J.-D.}\ \bibnamefont {Zheng}}, \ and\
  \bibinfo {author} {\bibfnamefont {C.-H.}\ \bibnamefont {Lu}},\ }\bibfield
  {title} {\enquote {\bibinfo {title} {{R}oles of {O}xygen {V}acancy in the
  {A}dsorption {P}roperties of \ce{CO} and \ce{NO} on \ce{Cu2O}(111) surface:
  {R}esults of a {F}irst-{P}rinciples {S}tudy},}\ }\href {\doibase
  https://doi.org/10.1016/j.apsusc.2008.09.005} {\bibfield  {journal} {\bibinfo
   {journal} {Appl. Surf. Sci.}\ }\textbf {\bibinfo {volume} {255}},\ \bibinfo
  {pages} {3141--3148} (\bibinfo {year} {2008})}\BibitemShut {NoStop}%
\bibitem [{\citenamefont {Bendavid}\ and\ \citenamefont
  {Carter}(2013)}]{Bendavid2013JPCB}%
  \BibitemOpen
  \bibfield  {author} {\bibinfo {author} {\bibfnamefont {L.~Isseroff}\
  \bibnamefont {Bendavid}}\ and\ \bibinfo {author} {\bibfnamefont {E.~A.}\
  \bibnamefont {Carter}},\ }\bibfield  {title} {\enquote {\bibinfo {title}
  {{F}irst-{P}rinciples {P}redictions of the {S}tructure, {S}tability, and
  {P}hotocatalytic {P}otential of \ce{Cu2O} surfaces},}\ }\href {\doibase
  10.1021/jp406454c} {\bibfield  {journal} {\bibinfo  {journal} {J. Phys. Chem.
  B}\ }\textbf {\bibinfo {volume} {117}},\ \bibinfo {pages} {15750--15760}
  (\bibinfo {year} {2013})}\BibitemShut {NoStop}%
\bibitem [{\citenamefont {Zhang}\ \emph {et~al.}(2018)\citenamefont {Zhang},
  \citenamefont {Li}, \citenamefont {Frazer}, \citenamefont {Chang},
  \citenamefont {Poeppelmeier}, \citenamefont {Chan},\ and\ \citenamefont
  {Guest}}]{Zhang2018}%
  \BibitemOpen
  \bibfield  {author} {\bibinfo {author} {\bibfnamefont {R.}~\bibnamefont
  {Zhang}}, \bibinfo {author} {\bibfnamefont {L.}~\bibnamefont {Li}}, \bibinfo
  {author} {\bibfnamefont {L.}~\bibnamefont {Frazer}}, \bibinfo {author}
  {\bibfnamefont {K.~B.}\ \bibnamefont {Chang}}, \bibinfo {author}
  {\bibfnamefont {K.~R.}\ \bibnamefont {Poeppelmeier}}, \bibinfo {author}
  {\bibfnamefont {M.~K.~Y.}\ \bibnamefont {Chan}}, \ and\ \bibinfo {author}
  {\bibfnamefont {J.~R.}\ \bibnamefont {Guest}},\ }\bibfield  {title} {\enquote
  {\bibinfo {title} {{A}tomistic {D}etermination of the {S}urface {S}tructure
  of \ce{Cu2O}(111): {E}xperiment and {T}heory},}\ }\href {\doibase
  10.1039/C8CP06023A} {\bibfield  {journal} {\bibinfo  {journal} {Phys. Chem.
  Chem. Phys.}\ }\textbf {\bibinfo {volume} {20}},\ \bibinfo {pages}
  {27456--27463} (\bibinfo {year} {2018})}\BibitemShut {NoStop}%
\bibitem [{\citenamefont {Yu}\ \emph {et~al.}(2018)\citenamefont {Yu},
  \citenamefont {Zhao}, \citenamefont {Zhang},\ and\ \citenamefont
  {Liu}}]{Yu2018}%
  \BibitemOpen
  \bibfield  {author} {\bibinfo {author} {\bibfnamefont {X.}~\bibnamefont
  {Yu}}, \bibinfo {author} {\bibfnamefont {C.}~\bibnamefont {Zhao}}, \bibinfo
  {author} {\bibfnamefont {T.}~\bibnamefont {Zhang}}, \ and\ \bibinfo {author}
  {\bibfnamefont {Z.}~\bibnamefont {Liu}},\ }\bibfield  {title} {\enquote
  {\bibinfo {title} {{M}olecular and {D}issociative \ce{O2} adsorption on the
  \ce{Cu2O}(111) {S}urface},}\ }\href {\doibase 10.1039/C8CP03035A} {\bibfield
  {journal} {\bibinfo  {journal} {Phys. Chem. Chem. Phys.}\ }\textbf {\bibinfo
  {volume} {20}},\ \bibinfo {pages} {20352--20362} (\bibinfo {year}
  {2018})}\BibitemShut {NoStop}%
\bibitem [{\citenamefont {Gloystein}\ \emph {et~al.}(2020)\citenamefont
  {Gloystein}, \citenamefont {Nilius}, \citenamefont {Goniakowski},\ and\
  \citenamefont {Noguera}}]{Gloystein2020}%
  \BibitemOpen
  \bibfield  {author} {\bibinfo {author} {\bibfnamefont {A.}~\bibnamefont
  {Gloystein}}, \bibinfo {author} {\bibfnamefont {N.}~\bibnamefont {Nilius}},
  \bibinfo {author} {\bibfnamefont {J.}~\bibnamefont {Goniakowski}}, \ and\
  \bibinfo {author} {\bibfnamefont {C.}~\bibnamefont {Noguera}},\ }\bibfield
  {title} {\enquote {\bibinfo {title} {{N}anopyramidal {R}econstruction of
  \ce{Cu2O}(111): {A} {L}ong-{S}tanding {S}urface {P}uzzle {S}olved by {STM}
  and {DFT}},}\ }\href {\doibase 10.1021/acs.jpcc.0c09330} {\bibfield
  {journal} {\bibinfo  {journal} {J. Phys. Chem.C}\ }\textbf {\bibinfo {volume}
  {124}},\ \bibinfo {pages} {26937--26943} (\bibinfo {year}
  {2020})}\BibitemShut {NoStop}%
\bibitem [{\citenamefont {Koirala}\ \emph {et~al.}(2014)\citenamefont
  {Koirala}, \citenamefont {Takahata}, \citenamefont {Hazama}, \citenamefont
  {Naka},\ and\ \citenamefont {Tanaka}}]{KOIRALA201465}%
  \BibitemOpen
  \bibfield  {author} {\bibinfo {author} {\bibfnamefont {S.}~\bibnamefont
  {Koirala}}, \bibinfo {author} {\bibfnamefont {M.}~\bibnamefont {Takahata}},
  \bibinfo {author} {\bibfnamefont {Y.}~\bibnamefont {Hazama}}, \bibinfo
  {author} {\bibfnamefont {N.}~\bibnamefont {Naka}}, \ and\ \bibinfo {author}
  {\bibfnamefont {K.}~\bibnamefont {Tanaka}},\ }\bibfield  {title} {\enquote
  {\bibinfo {title} {{R}elaxation of {L}ocalized {E}xcitons by {P}honon
  {E}mission at {O}xygen {V}acancies in \ce{Cu2O}},}\ }\href {\doibase
  https://doi.org/10.1016/j.jlumin.2014.06.027} {\bibfield  {journal} {\bibinfo
   {journal} {J. Lumin.}\ }\textbf {\bibinfo {volume} {155}},\ \bibinfo {pages}
  {65--69} (\bibinfo {year} {2014})}\BibitemShut {NoStop}%
\bibitem [{\citenamefont {Nolan}\ and\ \citenamefont
  {Elliott}(2006)}]{Nolan2006}%
  \BibitemOpen
  \bibfield  {author} {\bibinfo {author} {\bibfnamefont {M.}~\bibnamefont
  {Nolan}}\ and\ \bibinfo {author} {\bibfnamefont {S.~D.}\ \bibnamefont
  {Elliott}},\ }\bibfield  {title} {\enquote {\bibinfo {title} {{T}he p-type
  {C}onduction {M}echanism in \ce{Cu2O}: a {F}irst {P}rinciples {S}tudy},}\
  }\href {\doibase 10.1039/B611969G} {\bibfield  {journal} {\bibinfo  {journal}
  {Phys. Chem. Chem. Phys.}\ }\textbf {\bibinfo {volume} {8}},\ \bibinfo
  {pages} {5350--5358} (\bibinfo {year} {2006})}\BibitemShut {NoStop}%
\bibitem [{\citenamefont {Raebiger}\ \emph {et~al.}(2007)\citenamefont
  {Raebiger}, \citenamefont {Lany},\ and\ \citenamefont
  {Zunger}}]{Raebiger2007}%
  \BibitemOpen
  \bibfield  {author} {\bibinfo {author} {\bibfnamefont {H.}~\bibnamefont
  {Raebiger}}, \bibinfo {author} {\bibfnamefont {S.}~\bibnamefont {Lany}}, \
  and\ \bibinfo {author} {\bibfnamefont {A.}~\bibnamefont {Zunger}},\
  }\bibfield  {title} {\enquote {\bibinfo {title} {{O}rigins of the
  \textit{p}-type {N}ature and {C}ation {D}eficiency in \ce{Cu2O} and {R}elated
  {M}aterials},}\ }\href {\doibase 10.1103/PhysRevB.76.045209} {\bibfield
  {journal} {\bibinfo  {journal} {Phys. Rev. B}\ }\textbf {\bibinfo {volume}
  {76}},\ \bibinfo {pages} {045209} (\bibinfo {year} {2007})}\BibitemShut
  {NoStop}%
\bibitem [{\citenamefont {Nolan}(2008)}]{Nolan008}%
  \BibitemOpen
  \bibfield  {author} {\bibinfo {author} {\bibfnamefont {M.}~\bibnamefont
  {Nolan}},\ }\bibfield  {title} {\enquote {\bibinfo {title} {{D}efects in
  \ce{Cu2O}, \ce{CuAlO2} and \ce{SrCu2O2} {T}ransparent {C}onducting
  {O}xides},}\ }\href {\doibase https://doi.org/10.1016/j.tsf.2008.04.020}
  {\bibfield  {journal} {\bibinfo  {journal} {Thin Solid Films}\ }\textbf
  {\bibinfo {volume} {516}},\ \bibinfo {pages} {8130--8135} (\bibinfo {year}
  {2008})}\BibitemShut {NoStop}%
\bibitem [{\citenamefont {Scanlon}\ and\ \citenamefont
  {Watson}(2010)}]{ScanlonJPCL2010}%
  \BibitemOpen
  \bibfield  {author} {\bibinfo {author} {\bibfnamefont {D.~O.}\ \bibnamefont
  {Scanlon}}\ and\ \bibinfo {author} {\bibfnamefont {G.~W}\ \bibnamefont
  {Watson}},\ }\bibfield  {title} {\enquote {\bibinfo {title} {{U}ndoped n-type
  \ce{Cu2O}: {F}act or {F}iction?}}\ }\href {\doibase 10.1021/jz100962n}
  {\bibfield  {journal} {\bibinfo  {journal} {J. Phys. Chem. Lett.}\ }\textbf
  {\bibinfo {volume} {1}},\ \bibinfo {pages} {2582--2585} (\bibinfo {year}
  {2010})}\BibitemShut {NoStop}%
\bibitem [{\citenamefont {Huang}(2016)}]{Huang2016}%
  \BibitemOpen
  \bibfield  {author} {\bibinfo {author} {\bibfnamefont {B.}~\bibnamefont
  {Huang}},\ }\bibfield  {title} {\enquote {\bibinfo {title} {{I}ntrinsic
  {D}eep {H}ole {T}rap {L}evels in \ce{Cu2O} with {S}elf-{C}onsistent
  {R}epulsive {C}oulomb {E}nergy},}\ }\href {\doibase
  https://doi.org/10.1016/j.ssc.2016.01.008} {\bibfield  {journal} {\bibinfo
  {journal} {Solid State Commun.}\ }\textbf {\bibinfo {volume} {230}},\
  \bibinfo {pages} {49--53} (\bibinfo {year} {2016})}\BibitemShut {NoStop}%
\bibitem [{\citenamefont {Scanlon}\ \emph {et~al.}(2009)\citenamefont
  {Scanlon}, \citenamefont {Morgan}, \citenamefont {Watson},\ and\
  \citenamefont {Walsh}}]{ScanlonPRL2009}%
  \BibitemOpen
  \bibfield  {author} {\bibinfo {author} {\bibfnamefont {D.~O.}\ \bibnamefont
  {Scanlon}}, \bibinfo {author} {\bibfnamefont {B.~J.}\ \bibnamefont {Morgan}},
  \bibinfo {author} {\bibfnamefont {G.~W.}\ \bibnamefont {Watson}}, \ and\
  \bibinfo {author} {\bibfnamefont {A.}~\bibnamefont {Walsh}},\ }\bibfield
  {title} {\enquote {\bibinfo {title} {{A}cceptor {L}evels in \textit{p}-{T}ype
  \ce{Cu2O}: {R}ationalizing {T}heory and {E}xperiment},}\ }\href {\doibase
  10.1103/PhysRevLett.103.096405} {\bibfield  {journal} {\bibinfo  {journal}
  {Phys. Rev. Lett.}\ }\textbf {\bibinfo {volume} {103}},\ \bibinfo {pages}
  {096405} (\bibinfo {year} {2009})}\BibitemShut {NoStop}%
\bibitem [{\citenamefont {Kr{\"o}ger}\ and\ \citenamefont
  {Vink}(1956)}]{KROGER1956307}%
  \BibitemOpen
  \bibfield  {author} {\bibinfo {author} {\bibfnamefont {F.~A.}\ \bibnamefont
  {Kr{\"o}ger}}\ and\ \bibinfo {author} {\bibfnamefont {H.~J.}\ \bibnamefont
  {Vink}},\ }\bibfield  {title} {\enquote {\bibinfo {title} {{R}elations
  between the {C}oncentrations of {I}mperfections in {C}rystalline {S}olids},}\
  }\href {\doibase 10.1016/S0081-1947(08)60135-6} {\bibfield  {journal}
  {\bibinfo  {journal} {Solid State Phys.}\ }\textbf {\bibinfo {volume} {3}},\
  \bibinfo {pages} {307--435} (\bibinfo {year} {1956})}\BibitemShut {NoStop}%
\bibitem [{\citenamefont {Alkauskas}\ \emph {et~al.}(2016)\citenamefont
  {Alkauskas}, \citenamefont {Dreyer}, \citenamefont {Lyons},\ and\
  \citenamefont {Van~de Walle}}]{AlkauskasPRB2016}%
  \BibitemOpen
  \bibfield  {author} {\bibinfo {author} {\bibfnamefont {A.}~\bibnamefont
  {Alkauskas}}, \bibinfo {author} {\bibfnamefont {C.~E.}\ \bibnamefont
  {Dreyer}}, \bibinfo {author} {\bibfnamefont {J.~L.}\ \bibnamefont {Lyons}}, \
  and\ \bibinfo {author} {\bibfnamefont {C.~G.}\ \bibnamefont {Van~de Walle}},\
  }\bibfield  {title} {\enquote {\bibinfo {title} {{R}ole of {E}xcited {S}tates
  in {S}hockley-{R}ead-{H}all {R}ecombination in {W}ide-{B}and-{G}ap
  {S}emiconductors},}\ }\href {\doibase 10.1103/PhysRevB.93.201304} {\bibfield
  {journal} {\bibinfo  {journal} {Phys. Rev. B}\ }\textbf {\bibinfo {volume}
  {93}},\ \bibinfo {pages} {201304} (\bibinfo {year} {2016})}\BibitemShut
  {NoStop}%
\bibitem [{\citenamefont {Kresse}\ and\ \citenamefont
  {Hafner}(1993)}]{Kresse199347}%
  \BibitemOpen
  \bibfield  {author} {\bibinfo {author} {\bibfnamefont {G.}~\bibnamefont
  {Kresse}}\ and\ \bibinfo {author} {\bibfnamefont {J.}~\bibnamefont
  {Hafner}},\ }\bibfield  {title} {\enquote {\bibinfo {title} {{A}b {I}nitio
  {M}olecular {D}ynamics for {L}iquid {M}etals},}\ }\href {\doibase
  10.1103/PhysRevB.47.558} {\bibfield  {journal} {\bibinfo  {journal} {Phys.
  Rev. B}\ }\textbf {\bibinfo {volume} {47}},\ \bibinfo {pages} {558--561}
  (\bibinfo {year} {1993})}\BibitemShut {NoStop}%
\bibitem [{\citenamefont {Kresse}\ and\ \citenamefont
  {Hafner}(1994)}]{Kresse1994}%
  \BibitemOpen
  \bibfield  {author} {\bibinfo {author} {\bibfnamefont {G.}~\bibnamefont
  {Kresse}}\ and\ \bibinfo {author} {\bibfnamefont {J.}~\bibnamefont
  {Hafner}},\ }\bibfield  {title} {\enquote {\bibinfo {title} {{A}b {I}nitio
  {M}olecular-{D}ynamics {S}imulation of the
  {L}iquid-{M}etal--{A}morphous-{S}emiconductor {T}ransition in {G}ermanium},}\
  }\href {\doibase 10.1103/PhysRevB.49.14251} {\bibfield  {journal} {\bibinfo
  {journal} {Phys. Rev. B}\ }\textbf {\bibinfo {volume} {49}},\ \bibinfo
  {pages} {14251--14269} (\bibinfo {year} {1994})}\BibitemShut {NoStop}%
\bibitem [{\citenamefont {Kresse}\ and\ \citenamefont
  {Furthm{\"u}ller}(1996{\natexlab{a}})}]{KRESSE199615}%
  \BibitemOpen
  \bibfield  {author} {\bibinfo {author} {\bibfnamefont {G.}~\bibnamefont
  {Kresse}}\ and\ \bibinfo {author} {\bibfnamefont {J.}~\bibnamefont
  {Furthm{\"u}ller}},\ }\bibfield  {title} {\enquote {\bibinfo {title}
  {{E}fficiency of ab-initio {T}otal {E}nergy {C}alculations for {M}etals and
  {S}emiconductors using a {P}lane-{W}ave {B}asis {S}et},}\ }\href {\doibase
  https://doi.org/10.1016/0927-0256(96)00008-0} {\bibfield  {journal} {\bibinfo
   {journal} {Comput. Mater. Sci.}\ }\textbf {\bibinfo {volume} {6}},\ \bibinfo
  {pages} {15--50} (\bibinfo {year} {1996}{\natexlab{a}})}\BibitemShut
  {NoStop}%
\bibitem [{\citenamefont {Kresse}\ and\ \citenamefont
  {Furthm{\"u}ller}(1996{\natexlab{b}})}]{Kresse199654}%
  \BibitemOpen
  \bibfield  {author} {\bibinfo {author} {\bibfnamefont {G.}~\bibnamefont
  {Kresse}}\ and\ \bibinfo {author} {\bibfnamefont {J.}~\bibnamefont
  {Furthm{\"u}ller}},\ }\bibfield  {title} {\enquote {\bibinfo {title}
  {{E}fficient {I}terative {S}chemes for {A}b {I}nitio {T}otal-{E}nergy
  {C}alculations using a {P}lane-{W}ave {B}asis {S}et},}\ }\href {\doibase
  10.1103/PhysRevB.54.11169} {\bibfield  {journal} {\bibinfo  {journal} {Phys.
  Rev. B}\ }\textbf {\bibinfo {volume} {54}},\ \bibinfo {pages} {11169--11186}
  (\bibinfo {year} {1996}{\natexlab{b}})}\BibitemShut {NoStop}%
\bibitem [{\citenamefont {Heyd}\ \emph {et~al.}(2003)\citenamefont {Heyd},
  \citenamefont {Scuseria},\ and\ \citenamefont {Ernzerhof}}]{HeydJCP2003}%
  \BibitemOpen
  \bibfield  {author} {\bibinfo {author} {\bibfnamefont {J.}~\bibnamefont
  {Heyd}}, \bibinfo {author} {\bibfnamefont {G.~E.}\ \bibnamefont {Scuseria}},
  \ and\ \bibinfo {author} {\bibfnamefont {M.}~\bibnamefont {Ernzerhof}},\
  }\bibfield  {title} {\enquote {\bibinfo {title} {{H}ybrid {F}unctionals
  {B}ased on a {S}creened {C}oulomb {P}otential},}\ }\href {\doibase
  10.1063/1.1564060} {\bibfield  {journal} {\bibinfo  {journal} {J. Chem.
  Phys.}\ }\textbf {\bibinfo {volume} {118}},\ \bibinfo {pages} {8207--8215}
  (\bibinfo {year} {2003})}\BibitemShut {NoStop}%
\bibitem [{\citenamefont {Heyd}\ and\ \citenamefont
  {Scuseria}(2004)}]{HeydJCP2004}%
  \BibitemOpen
  \bibfield  {author} {\bibinfo {author} {\bibfnamefont {J.}~\bibnamefont
  {Heyd}}\ and\ \bibinfo {author} {\bibfnamefont {G.~E.}\ \bibnamefont
  {Scuseria}},\ }\bibfield  {title} {\enquote {\bibinfo {title} {{E}fficient
  {H}ybrid {D}ensity {F}unctional {C}alculations in {S}olids: {A}ssessment of
  the {H}eyd\textendash {S}cuseria\textendash {E}rnzerhof screened {C}oulomb
  {H}ybrid {F}unctional},}\ }\href {\doibase 10.1063/1.1760074} {\bibfield
  {journal} {\bibinfo  {journal} {J. Chem. Phys.}\ }\textbf {\bibinfo {volume}
  {121}},\ \bibinfo {pages} {1187--1192} (\bibinfo {year} {2004})}\BibitemShut
  {NoStop}%
\bibitem [{\citenamefont {Bl{\"o}chl}(1994)}]{BlochlPRB1994}%
  \BibitemOpen
  \bibfield  {author} {\bibinfo {author} {\bibfnamefont {P.~E.}\ \bibnamefont
  {Bl{\"o}chl}},\ }\bibfield  {title} {\enquote {\bibinfo {title} {{P}rojector
  {A}ugmented-{W}ave {M}ethod},}\ }\href {\doibase 10.1103/PhysRevB.50.17953}
  {\bibfield  {journal} {\bibinfo  {journal} {Phys. Rev. B}\ }\textbf {\bibinfo
  {volume} {50}},\ \bibinfo {pages} {17953--17979} (\bibinfo {year}
  {1994})}\BibitemShut {NoStop}%
\bibitem [{\citenamefont {Kresse}\ and\ \citenamefont
  {Joubert}(1999)}]{KressePRB1999}%
  \BibitemOpen
  \bibfield  {author} {\bibinfo {author} {\bibfnamefont {G.}~\bibnamefont
  {Kresse}}\ and\ \bibinfo {author} {\bibfnamefont {D.}~\bibnamefont
  {Joubert}},\ }\bibfield  {title} {\enquote {\bibinfo {title} {{F}rom
  {U}ltrasoft {P}seudopotentials to the {P}rojector {A}ugmented-{W}ave
  {M}ethod},}\ }\href {\doibase 10.1103/PhysRevB.59.1758} {\bibfield  {journal}
  {\bibinfo  {journal} {Phys. Rev. B}\ }\textbf {\bibinfo {volume} {59}},\
  \bibinfo {pages} {1758--1775} (\bibinfo {year} {1999})}\BibitemShut {NoStop}%
\bibitem [{\citenamefont {Freysoldt}\ \emph {et~al.}(2014)\citenamefont
  {Freysoldt}, \citenamefont {Grabowski}, \citenamefont {Hickel}, \citenamefont
  {Neugebauer}, \citenamefont {Kresse}, \citenamefont {Janotti},\ and\
  \citenamefont {Van~de Walle}}]{freysoldt2014first}%
  \BibitemOpen
  \bibfield  {author} {\bibinfo {author} {\bibfnamefont {C.}~\bibnamefont
  {Freysoldt}}, \bibinfo {author} {\bibfnamefont {B.}~\bibnamefont
  {Grabowski}}, \bibinfo {author} {\bibfnamefont {T.}~\bibnamefont {Hickel}},
  \bibinfo {author} {\bibfnamefont {J.}~\bibnamefont {Neugebauer}}, \bibinfo
  {author} {\bibfnamefont {G.}~\bibnamefont {Kresse}}, \bibinfo {author}
  {\bibfnamefont {A.}~\bibnamefont {Janotti}}, \ and\ \bibinfo {author}
  {\bibfnamefont {C.~G.}\ \bibnamefont {Van~de Walle}},\ }\bibfield  {title}
  {\enquote {\bibinfo {title} {{F}irst-{P}rinciples {C}alculations for {P}oint
  {D}efects in {S}olids},}\ }\href {\doibase 10.1103/RevModPhys.86.253}
  {\bibfield  {journal} {\bibinfo  {journal} {Rev. Mod. Phys.}\ }\textbf
  {\bibinfo {volume} {86}},\ \bibinfo {pages} {253} (\bibinfo {year}
  {2014})}\BibitemShut {NoStop}%
\bibitem [{\citenamefont {Lany}\ and\ \citenamefont {Zunger}(2008)}]{lany2008}%
  \BibitemOpen
  \bibfield  {author} {\bibinfo {author} {\bibfnamefont {S.}~\bibnamefont
  {Lany}}\ and\ \bibinfo {author} {\bibfnamefont {A.}~\bibnamefont {Zunger}},\
  }\bibfield  {title} {\enquote {\bibinfo {title} {{A}ssessment of {C}orrection
  {M}ethods for the {B}and-{G}ap {P}roblem and for {F}inite-{S}ize {E}ffects in
  {S}upercell {D}efect {C}alculations: {C}ase {S}tudies for zno and
  {G}a{A}s},}\ }\href {\doibase 10.1103/PhysRevB.78.235104} {\bibfield
  {journal} {\bibinfo  {journal} {Phys. Rev. B}\ }\textbf {\bibinfo {volume}
  {78}},\ \bibinfo {pages} {235104} (\bibinfo {year} {2008})}\BibitemShut
  {NoStop}%
\bibitem [{\citenamefont {Huang}\ \emph {et~al.}(1950)\citenamefont {Huang},
  \citenamefont {Rhys},\ and\ \citenamefont {Mott}}]{Huang1950}%
  \BibitemOpen
  \bibfield  {author} {\bibinfo {author} {\bibfnamefont {K.}~\bibnamefont
  {Huang}}, \bibinfo {author} {\bibfnamefont {A.}~\bibnamefont {Rhys}}, \ and\
  \bibinfo {author} {\bibfnamefont {N.~F.}\ \bibnamefont {Mott}},\ }\bibfield
  {title} {\enquote {\bibinfo {title} {{T}heory of {L}ight {A}bsorption and
  {N}on-{R}adiative {T}ransitions in \textit{F}-centres},}\ }\href {\doibase
  10.1098/rspa.1950.0184} {\bibfield  {journal} {\bibinfo  {journal} {Proc. R.
  Soc. Lond. Series A. Mathematical and Physical Sciences}\ }\textbf {\bibinfo
  {volume} {204}},\ \bibinfo {pages} {406--423} (\bibinfo {year}
  {1950})}\BibitemShut {NoStop}%
\bibitem [{\citenamefont {Henry}\ and\ \citenamefont
  {Lang}(1977)}]{HenryLang1977}%
  \BibitemOpen
  \bibfield  {author} {\bibinfo {author} {\bibfnamefont {C.~H.}\ \bibnamefont
  {Henry}}\ and\ \bibinfo {author} {\bibfnamefont {D.~V.}\ \bibnamefont
  {Lang}},\ }\bibfield  {title} {\enquote {\bibinfo {title} {{N}onradiative
  {C}apture and {R}ecombination by {M}ultiphonon {E}mission in {G}a{A}s and
  {G}a{P}},}\ }\href {\doibase 10.1103/PhysRevB.15.989} {\bibfield  {journal}
  {\bibinfo  {journal} {Phys. Rev. B}\ }\textbf {\bibinfo {volume} {15}},\
  \bibinfo {pages} {989--1016} (\bibinfo {year} {1977})}\BibitemShut {NoStop}%
\bibitem [{\citenamefont {Alkauskas}\ \emph {et~al.}(2014)\citenamefont
  {Alkauskas}, \citenamefont {Yan},\ and\ \citenamefont {Van~de
  Walle}}]{AlkauskasPRB2014}%
  \BibitemOpen
  \bibfield  {author} {\bibinfo {author} {\bibfnamefont {A.}~\bibnamefont
  {Alkauskas}}, \bibinfo {author} {\bibfnamefont {Q.}~\bibnamefont {Yan}}, \
  and\ \bibinfo {author} {\bibfnamefont {C.~G.}\ \bibnamefont {Van~de Walle}},\
  }\bibfield  {title} {\enquote {\bibinfo {title} {{F}irst-{P}rinciples
  {T}heory of {N}onradiative {C}arrier {C}apture via {M}ultiphonon
  {E}mission},}\ }\href {\doibase 10.1103/PhysRevB.90.075202} {\bibfield
  {journal} {\bibinfo  {journal} {Phys. Rev. B}\ }\textbf {\bibinfo {volume}
  {90}},\ \bibinfo {pages} {075202} (\bibinfo {year} {2014})}\BibitemShut
  {NoStop}%
\bibitem [{\citenamefont {Huang}(1981)}]{Huang1981}%
  \BibitemOpen
  \bibfield  {author} {\bibinfo {author} {\bibfnamefont {H.}~\bibnamefont
  {Huang}},\ }\bibfield  {title} {\enquote {\bibinfo {title} {{A}diabatic
  {A}pproximation {T}heory and {S}tatic {C}oupling {T}heory of {N}onradiative
  {T}ransition},}\ }\href@noop {} {\bibfield  {journal} {\bibinfo  {journal}
  {Sci. Sin.}\ }\textbf {\bibinfo {volume} {24}},\ \bibinfo {pages} {27--34}
  (\bibinfo {year} {1981})}\BibitemShut {NoStop}%
\bibitem [{\citenamefont {Bystrom}\ \emph {et~al.}(2019)\citenamefont
  {Bystrom}, \citenamefont {Broberg}, \citenamefont {Dwaraknath}, \citenamefont
  {Persson},\ and\ \citenamefont {Asta}}]{pawpyseed}%
  \BibitemOpen
  \bibfield  {author} {\bibinfo {author} {\bibfnamefont {K.}~\bibnamefont
  {Bystrom}}, \bibinfo {author} {\bibfnamefont {D.}~\bibnamefont {Broberg}},
  \bibinfo {author} {\bibfnamefont {S.}~\bibnamefont {Dwaraknath}}, \bibinfo
  {author} {\bibfnamefont {K.~A.}\ \bibnamefont {Persson}}, \ and\ \bibinfo
  {author} {\bibfnamefont {M.}~\bibnamefont {Asta}},\ }\bibfield  {title}
  {\enquote {\bibinfo {title} {{P}awpyseed: {P}erturbation-{E}xtrapolation
  {B}and {S}hifting {C}orrections for {P}oint {D}efect {C}alculations},}\
  }\href@noop {} {\  (\bibinfo {year} {2019})},\ \Eprint
  {http://arxiv.org/abs/1904.11572} {arXiv:1904.11572 [cond-mat.mtrl-sci]}
  \BibitemShut {NoStop}%
\bibitem [{\citenamefont {Kim}\ \emph {et~al.}(2019{\natexlab{a}})\citenamefont
  {Kim}, \citenamefont {Park}, \citenamefont {Hood},\ and\ \citenamefont
  {Walsh}}]{KimJmatChemA2019}%
  \BibitemOpen
  \bibfield  {author} {\bibinfo {author} {\bibfnamefont {S.}~\bibnamefont
  {Kim}}, \bibinfo {author} {\bibfnamefont {J.-S.}\ \bibnamefont {Park}},
  \bibinfo {author} {\bibfnamefont {S.~N.}\ \bibnamefont {Hood}}, \ and\
  \bibinfo {author} {\bibfnamefont {A.}~\bibnamefont {Walsh}},\ }\bibfield
  {title} {\enquote {\bibinfo {title} {{L}one-{P}air {E}ffect on {C}arrier
  {C}apture in \ce{Cu2ZnSnS4} {S}olar {C}ells},}\ }\href {\doibase
  10.1039/C8TA10130B} {\bibfield  {journal} {\bibinfo  {journal} {J. Mater.
  Chem. A}\ }\textbf {\bibinfo {volume} {7}},\ \bibinfo {pages} {2686--2693}
  (\bibinfo {year} {2019}{\natexlab{a}})}\BibitemShut {NoStop}%
\bibitem [{\citenamefont {Kim}\ \emph {et~al.}(2020)\citenamefont {Kim},
  \citenamefont {M{\'a}rquez}, \citenamefont {Unold},\ and\ \citenamefont
  {Walsh}}]{KimRSC2020}%
  \BibitemOpen
  \bibfield  {author} {\bibinfo {author} {\bibfnamefont {S.}~\bibnamefont
  {Kim}}, \bibinfo {author} {\bibfnamefont {J.~A.}\ \bibnamefont
  {M{\'a}rquez}}, \bibinfo {author} {\bibfnamefont {T.}~\bibnamefont {Unold}},
  \ and\ \bibinfo {author} {\bibfnamefont {A.}~\bibnamefont {Walsh}},\
  }\bibfield  {title} {\enquote {\bibinfo {title} {{U}pper {L}imit to the
  {P}hotovoltaic {E}fficiency of {I}mperfect {C}rystals from {F}irst
  {P}rinciples},}\ }\href {\doibase 10.1039/D0EE00291G} {\bibfield  {journal}
  {\bibinfo  {journal} {Energy Environ. Sci.}\ }\textbf {\bibinfo {volume}
  {13}},\ \bibinfo {pages} {1481--1491} (\bibinfo {year} {2020})}\BibitemShut
  {NoStop}%
\bibitem [{\citenamefont {Fonari}\ and\ \citenamefont
  {Sutton}(2012)}]{fonari2012effective}%
  \BibitemOpen
  \bibfield  {author} {\bibinfo {author} {\bibfnamefont {A.}~\bibnamefont
  {Fonari}}\ and\ \bibinfo {author} {\bibfnamefont {C.}~\bibnamefont
  {Sutton}},\ }\bibfield  {title} {\enquote {\bibinfo {title} {{E}ffective
  {M}ass {C}alculator},}\ }\href@noop {} {\  (\bibinfo {year}
  {2012})}\BibitemShut {NoStop}%
\bibitem [{\citenamefont {Kim}\ \emph {et~al.}(2019{\natexlab{b}})\citenamefont
  {Kim}, \citenamefont {Hood},\ and\ \citenamefont {Walsh}}]{carriercapture}%
  \BibitemOpen
  \bibfield  {author} {\bibinfo {author} {\bibfnamefont {S.}~\bibnamefont
  {Kim}}, \bibinfo {author} {\bibfnamefont {N.~S.}\ \bibnamefont {Hood}}, \
  and\ \bibinfo {author} {\bibfnamefont {A.}~\bibnamefont {Walsh}},\
  }\href@noop {} {\enquote {\bibinfo {title} {{C}arriercapture},}\ }\bibinfo
  {howpublished} {https://github.com/WMD-group/CarrierCapture.jl} (\bibinfo
  {year} {2019}{\natexlab{b}})\BibitemShut {NoStop}%
\end{thebibliography}%


\begin{thebibliography}{35}%
\makeatletter
\providecommand \@ifxundefined [1]{%
 \@ifx{#1\undefined}
}%
\providecommand \@ifnum [1]{%
 \ifnum #1\expandafter \@firstoftwo
 \else \expandafter \@secondoftwo
 \fi
}%
\providecommand \@ifx [1]{%
 \ifx #1\expandafter \@firstoftwo
 \else \expandafter \@secondoftwo
 \fi
}%
\providecommand \natexlab [1]{#1}%
\providecommand \enquote  [1]{``#1''}%
\providecommand \bibnamefont  [1]{#1}%
\providecommand \bibfnamefont [1]{#1}%
\providecommand \citenamefont [1]{#1}%
\providecommand \href@noop [0]{\@secondoftwo}%
\providecommand \href [0]{\begingroup \@sanitize@url \@href}%
\providecommand \@href[1]{\@@startlink{#1}\@@href}%
\providecommand \@@href[1]{\endgroup#1\@@endlink}%
\providecommand \@sanitize@url [0]{\catcode `\\12\catcode `\$12\catcode
  `\&12\catcode `\#12\catcode `\^12\catcode `\_12\catcode `\%12\relax}%
\providecommand \@@startlink[1]{}%
\providecommand \@@endlink[0]{}%
\providecommand \url  [0]{\begingroup\@sanitize@url \@url }%
\providecommand \@url [1]{\endgroup\@href {#1}{\urlprefix }}%
\providecommand \urlprefix  [0]{URL }%
\providecommand \Eprint [0]{\href }%
\providecommand \doibase [0]{http://dx.doi.org/}%
\providecommand \selectlanguage [0]{\@gobble}%
\providecommand \bibinfo  [0]{\@secondoftwo}%
\providecommand \bibfield  [0]{\@secondoftwo}%
\providecommand \translation [1]{[#1]}%
\providecommand \BibitemOpen [0]{}%
\providecommand \bibitemStop [0]{}%
\providecommand \bibitemNoStop [0]{.\EOS\space}%
\providecommand \EOS [0]{\spacefactor3000\relax}%
\providecommand \BibitemShut  [1]{\csname bibitem#1\endcsname}%
\let\auto@bib@innerbib\@empty
\bibitem [{\citenamefont {Monkhorst}\ and\ \citenamefont
  {Pack}(1976)}]{SI_monkhorst1976special}%
  \BibitemOpen
  \bibfield  {author} {\bibinfo {author} {\bibfnamefont {H.~J.}\ \bibnamefont
  {Monkhorst}}\ and\ \bibinfo {author} {\bibfnamefont {J.~D.}\ \bibnamefont
  {Pack}},\ }\href {\doibase 10.1103/PhysRevB.13.5188} {\bibfield  {journal}
  {\bibinfo  {journal} {Phys. Rev. B}\ }\textbf {\bibinfo {volume} {13}},\
  \bibinfo {pages} {5188} (\bibinfo {year} {1976})}\BibitemShut {NoStop}%
\bibitem [{\citenamefont {Makov}\ and\ \citenamefont
  {Payne}(1995)}]{SI_makov1995periodic}%
  \BibitemOpen
  \bibfield  {author} {\bibinfo {author} {\bibfnamefont {G.}~\bibnamefont
  {Makov}}\ and\ \bibinfo {author} {\bibfnamefont {M.~C.}\ \bibnamefont
  {Payne}},\ }\href {\doibase 10.1103/physrevb.51.4014} {\bibfield  {journal}
  {\bibinfo  {journal} {Phys. Rev. B}\ }\textbf {\bibinfo {volume} {51}},\
  \bibinfo {pages} {4014} (\bibinfo {year} {1995})}\BibitemShut {NoStop}%
\bibitem [{\citenamefont {Lany}\ and\ \citenamefont
  {Zunger}(2009)}]{SI_lany2009accurate}%
  \BibitemOpen
  \bibfield  {author} {\bibinfo {author} {\bibfnamefont {S.}~\bibnamefont
  {Lany}}\ and\ \bibinfo {author} {\bibfnamefont {A.}~\bibnamefont {Zunger}},\
  }\href {\doibase 10.1088/0965-0393/17/8/084002} {\bibfield  {journal}
  {\bibinfo  {journal} {Model. Simul. Mater. Sci. Eng.}\ }\textbf {\bibinfo
  {volume} {17}},\ \bibinfo {pages} {084002} (\bibinfo {year}
  {2009})}\BibitemShut {NoStop}%
\bibitem [{\citenamefont {Stolz}\ \emph {et~al.}(2018)\citenamefont {Stolz},
  \citenamefont {Sch{\"o}ne},\ and\ \citenamefont {Semkat}}]{SI_Stolz2018}%
  \BibitemOpen
  \bibfield  {author} {\bibinfo {author} {\bibfnamefont {H.}~\bibnamefont
  {Stolz}}, \bibinfo {author} {\bibfnamefont {F.}~\bibnamefont {Sch{\"o}ne}}, \
  and\ \bibinfo {author} {\bibfnamefont {D.}~\bibnamefont {Semkat}},\ }\href
  {\doibase 10.1088/1367-2630/aaa396} {\bibfield  {journal} {\bibinfo
  {journal} {New J. Phys.}\ }\textbf {\bibinfo {volume} {20}},\ \bibinfo
  {pages} {023019} (\bibinfo {year} {2018})}\BibitemShut {NoStop}%
\bibitem [{\citenamefont {Huang}\ \emph {et~al.}(1950)\citenamefont {Huang},
  \citenamefont {Rhys},\ and\ \citenamefont {Mott}}]{SI_Huang1950}%
  \BibitemOpen
  \bibfield  {author} {\bibinfo {author} {\bibfnamefont {K.}~\bibnamefont
  {Huang}}, \bibinfo {author} {\bibfnamefont {A.}~\bibnamefont {Rhys}}, \ and\
  \bibinfo {author} {\bibfnamefont {N.~F.}\ \bibnamefont {Mott}},\ }\href
  {\doibase 10.1098/rspa.1950.0184} {\bibfield  {journal} {\bibinfo  {journal}
  {Proc. R. Soc. Lond. Series A. Mathematical and Physical Sciences}\ }\textbf
  {\bibinfo {volume} {204}},\ \bibinfo {pages} {406} (\bibinfo {year}
  {1950})}\BibitemShut {NoStop}%
\bibitem [{\citenamefont {Henry}\ and\ \citenamefont
  {Lang}(1977)}]{SI_HenryLang1977}%
  \BibitemOpen
  \bibfield  {author} {\bibinfo {author} {\bibfnamefont {C.~H.}\ \bibnamefont
  {Henry}}\ and\ \bibinfo {author} {\bibfnamefont {D.~V.}\ \bibnamefont
  {Lang}},\ }\href {\doibase 10.1103/PhysRevB.15.989} {\bibfield  {journal}
  {\bibinfo  {journal} {Phys. Rev. B}\ }\textbf {\bibinfo {volume} {15}},\
  \bibinfo {pages} {989} (\bibinfo {year} {1977})}\BibitemShut {NoStop}%
\bibitem [{\citenamefont {Alkauskas}\ \emph {et~al.}(2014)\citenamefont
  {Alkauskas}, \citenamefont {Yan},\ and\ \citenamefont {Van~de
  Walle}}]{SI_AlkauskasPRB2014}%
  \BibitemOpen
  \bibfield  {author} {\bibinfo {author} {\bibfnamefont {A.}~\bibnamefont
  {Alkauskas}}, \bibinfo {author} {\bibfnamefont {Q.}~\bibnamefont {Yan}}, \
  and\ \bibinfo {author} {\bibfnamefont {C.~G.}\ \bibnamefont {Van~de Walle}},\
  }\href {\doibase 10.1103/PhysRevB.90.075202} {\bibfield  {journal} {\bibinfo
  {journal} {Phys. Rev. B}\ }\textbf {\bibinfo {volume} {90}},\ \bibinfo
  {pages} {075202} (\bibinfo {year} {2014})}\BibitemShut {NoStop}%
\bibitem [{\citenamefont {Alkauskas}\ \emph {et~al.}(2016)\citenamefont
  {Alkauskas}, \citenamefont {Dreyer}, \citenamefont {Lyons},\ and\
  \citenamefont {Van~de Walle}}]{SI_AlkauskasPRB2016}%
  \BibitemOpen
  \bibfield  {author} {\bibinfo {author} {\bibfnamefont {A.}~\bibnamefont
  {Alkauskas}}, \bibinfo {author} {\bibfnamefont {C.~E.}\ \bibnamefont
  {Dreyer}}, \bibinfo {author} {\bibfnamefont {J.~L.}\ \bibnamefont {Lyons}}, \
  and\ \bibinfo {author} {\bibfnamefont {C.~G.}\ \bibnamefont {Van~de Walle}},\
  }\href {\doibase 10.1103/PhysRevB.93.201304} {\bibfield  {journal} {\bibinfo
  {journal} {Phys. Rev. B}\ }\textbf {\bibinfo {volume} {93}},\ \bibinfo
  {pages} {201304} (\bibinfo {year} {2016})}\BibitemShut {NoStop}%
\bibitem [{\citenamefont {Huang}(1981)}]{SI_Huang1981}%
  \BibitemOpen
  \bibfield  {author} {\bibinfo {author} {\bibfnamefont {H.}~\bibnamefont
  {Huang}},\ }\href@noop {} {\bibfield  {journal} {\bibinfo  {journal} {Sci.
  Sin.}\ }\textbf {\bibinfo {volume} {24}},\ \bibinfo {pages} {27} (\bibinfo
  {year} {1981})}\BibitemShut {NoStop}%
\bibitem [{\citenamefont {Kim}\ \emph {et~al.}(2019)\citenamefont {Kim},
  \citenamefont {Park}, \citenamefont {Hood},\ and\ \citenamefont
  {Walsh}}]{SI_KimJmatChemA2019}%
  \BibitemOpen
  \bibfield  {author} {\bibinfo {author} {\bibfnamefont {S.}~\bibnamefont
  {Kim}}, \bibinfo {author} {\bibfnamefont {J.-S.}\ \bibnamefont {Park}},
  \bibinfo {author} {\bibfnamefont {S.~N.}\ \bibnamefont {Hood}}, \ and\
  \bibinfo {author} {\bibfnamefont {A.}~\bibnamefont {Walsh}},\ }\href
  {\doibase 10.1039/C8TA10130B} {\bibfield  {journal} {\bibinfo  {journal} {J.
  Mater. Chem. A}\ }\textbf {\bibinfo {volume} {7}},\ \bibinfo {pages} {2686}
  (\bibinfo {year} {2019})}\BibitemShut {NoStop}%
\bibitem [{\citenamefont {Kim}\ \emph {et~al.}(2020)\citenamefont {Kim},
  \citenamefont {M{\'a}rquez}, \citenamefont {Unold},\ and\ \citenamefont
  {Walsh}}]{SI_KimRSC2020}%
  \BibitemOpen
  \bibfield  {author} {\bibinfo {author} {\bibfnamefont {S.}~\bibnamefont
  {Kim}}, \bibinfo {author} {\bibfnamefont {J.~A.}\ \bibnamefont
  {M{\'a}rquez}}, \bibinfo {author} {\bibfnamefont {T.}~\bibnamefont {Unold}},
  \ and\ \bibinfo {author} {\bibfnamefont {A.}~\bibnamefont {Walsh}},\ }\href
  {\doibase 10.1039/D0EE00291G} {\bibfield  {journal} {\bibinfo  {journal}
  {Energy Environ. Sci.}\ }\textbf {\bibinfo {volume} {13}},\ \bibinfo {pages}
  {1481} (\bibinfo {year} {2020})}\BibitemShut {NoStop}%
\bibitem [{\citenamefont {Scanlon}\ \emph
  {et~al.}(2009{\natexlab{a}})\citenamefont {Scanlon}, \citenamefont {Morgan},
  \citenamefont {Watson},\ and\ \citenamefont {Walsh}}]{SI_ScanlonPRL2009}%
  \BibitemOpen
  \bibfield  {author} {\bibinfo {author} {\bibfnamefont {D.~O.}\ \bibnamefont
  {Scanlon}}, \bibinfo {author} {\bibfnamefont {B.~J.}\ \bibnamefont {Morgan}},
  \bibinfo {author} {\bibfnamefont {G.~W.}\ \bibnamefont {Watson}}, \ and\
  \bibinfo {author} {\bibfnamefont {A.}~\bibnamefont {Walsh}},\ }\href
  {\doibase 10.1103/PhysRevLett.103.096405} {\bibfield  {journal} {\bibinfo
  {journal} {Phys. Rev. Lett.}\ }\textbf {\bibinfo {volume} {103}},\ \bibinfo
  {pages} {096405} (\bibinfo {year} {2009}{\natexlab{a}})}\BibitemShut
  {NoStop}%
\bibitem [{\citenamefont {Scanlon}\ and\ \citenamefont
  {Watson}(2010)}]{SI_ScanlonJPCL2010}%
  \BibitemOpen
  \bibfield  {author} {\bibinfo {author} {\bibfnamefont {D.~O.}\ \bibnamefont
  {Scanlon}}\ and\ \bibinfo {author} {\bibfnamefont {G.~W.}\ \bibnamefont
  {Watson}},\ }\href {\doibase 10.1021/jz100962n} {\bibfield  {journal}
  {\bibinfo  {journal} {J. Phys. Chem. Lett.}\ }\textbf {\bibinfo {volume}
  {1}},\ \bibinfo {pages} {2582} (\bibinfo {year} {2010})}\BibitemShut
  {NoStop}%
\bibitem [{\citenamefont {Heinemann}\ \emph {et~al.}(2013)\citenamefont
  {Heinemann}, \citenamefont {Eifert},\ and\ \citenamefont
  {Heiliger}}]{SI_Heinemann2013}%
  \BibitemOpen
  \bibfield  {author} {\bibinfo {author} {\bibfnamefont {M.}~\bibnamefont
  {Heinemann}}, \bibinfo {author} {\bibfnamefont {B.}~\bibnamefont {Eifert}}, \
  and\ \bibinfo {author} {\bibfnamefont {C.}~\bibnamefont {Heiliger}},\ }\href
  {\doibase 10.1103/PhysRevB.87.115111} {\bibfield  {journal} {\bibinfo
  {journal} {Phys. Rev. B}\ }\textbf {\bibinfo {volume} {87}},\ \bibinfo
  {pages} {115111} (\bibinfo {year} {2013})}\BibitemShut {NoStop}%
\bibitem [{\citenamefont {Nolan}\ and\ \citenamefont
  {Elliott}(2006)}]{SI_Nolan2006}%
  \BibitemOpen
  \bibfield  {author} {\bibinfo {author} {\bibfnamefont {M.}~\bibnamefont
  {Nolan}}\ and\ \bibinfo {author} {\bibfnamefont {S.~D.}\ \bibnamefont
  {Elliott}},\ }\href {\doibase 10.1039/B611969G} {\bibfield  {journal}
  {\bibinfo  {journal} {Phys. Chem. Chem. Phys.}\ }\textbf {\bibinfo {volume}
  {8}},\ \bibinfo {pages} {5350} (\bibinfo {year} {2006})}\BibitemShut
  {NoStop}%
\bibitem [{\citenamefont {Werner}\ and\ \citenamefont
  {Hochheimer}(1982)}]{SI_Werner1982}%
  \BibitemOpen
  \bibfield  {author} {\bibinfo {author} {\bibfnamefont {A.}~\bibnamefont
  {Werner}}\ and\ \bibinfo {author} {\bibfnamefont {H.~D.}\ \bibnamefont
  {Hochheimer}},\ }\href {\doibase 10.1103/PhysRevB.25.5929} {\bibfield
  {journal} {\bibinfo  {journal} {Phys. Rev. B}\ }\textbf {\bibinfo {volume}
  {25}},\ \bibinfo {pages} {5929} (\bibinfo {year} {1982})}\BibitemShut
  {NoStop}%
\bibitem [{SI_(2008)}]{SI_Lide2008}%
  \BibitemOpen
  \href {\doibase 10.1021/ja077011d} {\bibfield  {journal} {\bibinfo  {journal}
  {J. Am. Chem. Soc.}\ }\textbf {\bibinfo {volume} {130}},\ \bibinfo {pages}
  {382} (\bibinfo {year} {2008})}\BibitemShut {NoStop}%
\bibitem [{\citenamefont {Raebiger}\ \emph {et~al.}(2007)\citenamefont
  {Raebiger}, \citenamefont {Lany},\ and\ \citenamefont
  {Zunger}}]{SI_Raebiger2007}%
  \BibitemOpen
  \bibfield  {author} {\bibinfo {author} {\bibfnamefont {H.}~\bibnamefont
  {Raebiger}}, \bibinfo {author} {\bibfnamefont {S.}~\bibnamefont {Lany}}, \
  and\ \bibinfo {author} {\bibfnamefont {A.}~\bibnamefont {Zunger}},\ }\href
  {\doibase 10.1103/PhysRevB.76.045209} {\bibfield  {journal} {\bibinfo
  {journal} {Phys. Rev. B}\ }\textbf {\bibinfo {volume} {76}},\ \bibinfo
  {pages} {045209} (\bibinfo {year} {2007})}\BibitemShut {NoStop}%
\bibitem [{\citenamefont {Ohyama}\ \emph {et~al.}(1997)\citenamefont {Ohyama},
  \citenamefont {Ogawa},\ and\ \citenamefont {Nakata}}]{SI_Ohyama1997}%
  \BibitemOpen
  \bibfield  {author} {\bibinfo {author} {\bibfnamefont {T.}~\bibnamefont
  {Ohyama}}, \bibinfo {author} {\bibfnamefont {T.}~\bibnamefont {Ogawa}}, \
  and\ \bibinfo {author} {\bibfnamefont {H.}~\bibnamefont {Nakata}},\ }\href
  {\doibase 10.1103/PhysRevB.56.3871} {\bibfield  {journal} {\bibinfo
  {journal} {Phys. Rev. B}\ }\textbf {\bibinfo {volume} {56}},\ \bibinfo
  {pages} {3871} (\bibinfo {year} {1997})}\BibitemShut {NoStop}%
\bibitem [{\citenamefont {Nie}\ \emph {et~al.}(2002)\citenamefont {Nie},
  \citenamefont {Wei},\ and\ \citenamefont {Zhang}}]{SI_Nie2002}%
  \BibitemOpen
  \bibfield  {author} {\bibinfo {author} {\bibfnamefont {X.}~\bibnamefont
  {Nie}}, \bibinfo {author} {\bibfnamefont {S.-H.}\ \bibnamefont {Wei}}, \ and\
  \bibinfo {author} {\bibfnamefont {S.~B.}\ \bibnamefont {Zhang}},\ }\href
  {\doibase 10.1103/PhysRevB.65.075111} {\bibfield  {journal} {\bibinfo
  {journal} {Phys. Rev. B}\ }\textbf {\bibinfo {volume} {65}},\ \bibinfo
  {pages} {075111} (\bibinfo {year} {2002})}\BibitemShut {NoStop}%
\bibitem [{\citenamefont {Hodby}\ \emph {et~al.}(1976)\citenamefont {Hodby},
  \citenamefont {Jenkins}, \citenamefont {Schwab}, \citenamefont {Tamura},\
  and\ \citenamefont {Trivich}}]{SI_Hodby_1976}%
  \BibitemOpen
  \bibfield  {author} {\bibinfo {author} {\bibfnamefont {J.~W.}\ \bibnamefont
  {Hodby}}, \bibinfo {author} {\bibfnamefont {T.~E.}\ \bibnamefont {Jenkins}},
  \bibinfo {author} {\bibfnamefont {C.}~\bibnamefont {Schwab}}, \bibinfo
  {author} {\bibfnamefont {H.}~\bibnamefont {Tamura}}, \ and\ \bibinfo {author}
  {\bibfnamefont {D.}~\bibnamefont {Trivich}},\ }\href {\doibase
  10.1088/0022-3719/9/8/014} {\bibfield  {journal} {\bibinfo  {journal} {J.
  Phys. C. Solid State Phys.}\ }\textbf {\bibinfo {volume} {9}},\ \bibinfo
  {pages} {1429} (\bibinfo {year} {1976})}\BibitemShut {NoStop}%
\bibitem [{\citenamefont {Goltzene}\ \emph {et~al.}(1976)\citenamefont
  {Goltzene}, \citenamefont {Schwab},\ and\ \citenamefont
  {Wolf}}]{SI_GOLTZENE1976}%
  \BibitemOpen
  \bibfield  {author} {\bibinfo {author} {\bibfnamefont {A.}~\bibnamefont
  {Goltzene}}, \bibinfo {author} {\bibfnamefont {C.}~\bibnamefont {Schwab}}, \
  and\ \bibinfo {author} {\bibfnamefont {H.~C.}\ \bibnamefont {Wolf}},\ }\href
  {\doibase https://doi.org/10.1016/0038-1098(76)90394-X} {\bibfield  {journal}
  {\bibinfo  {journal} {Solid State Commun.}\ }\textbf {\bibinfo {volume}
  {18}},\ \bibinfo {pages} {1565} (\bibinfo {year} {1976})}\BibitemShut
  {NoStop}%
\bibitem [{\citenamefont {Bloem}(1958)}]{SI_Bloem1958}%
  \BibitemOpen
  \bibfield  {author} {\bibinfo {author} {\bibfnamefont {J.}~\bibnamefont
  {Bloem}},\ }\href@noop {} {\bibfield  {journal} {\bibinfo  {journal}
  {Phillips Res. Rep.}\ }\textbf {\bibinfo {volume} {31}},\ \bibinfo {pages}
  {167} (\bibinfo {year} {1958})}\BibitemShut {NoStop}%
\bibitem [{\citenamefont {Zouaghi}\ \emph {et~al.}(1969)\citenamefont
  {Zouaghi}, \citenamefont {Coret},\ and\ \citenamefont
  {Eymann}}]{SI_ZOUAGHI1969311}%
  \BibitemOpen
  \bibfield  {author} {\bibinfo {author} {\bibfnamefont {M.}~\bibnamefont
  {Zouaghi}}, \bibinfo {author} {\bibfnamefont {A.}~\bibnamefont {Coret}}, \
  and\ \bibinfo {author} {\bibfnamefont {J.~O.}\ \bibnamefont {Eymann}},\
  }\href {\doibase https://doi.org/10.1016/0038-1098(69)90408-6} {\bibfield
  {journal} {\bibinfo  {journal} {Solid State Commun.}\ }\textbf {\bibinfo
  {volume} {7}},\ \bibinfo {pages} {311} (\bibinfo {year} {1969})}\BibitemShut
  {NoStop}%
\bibitem [{\citenamefont {Gastev}\ \emph {et~al.}(1982)\citenamefont {Gastev},
  \citenamefont {Kaplyanskii},\ and\ \citenamefont
  {Sokolov}}]{SI_GASTEV1982389}%
  \BibitemOpen
  \bibfield  {author} {\bibinfo {author} {\bibfnamefont {S.~V.}\ \bibnamefont
  {Gastev}}, \bibinfo {author} {\bibfnamefont {A.~A.}\ \bibnamefont
  {Kaplyanskii}}, \ and\ \bibinfo {author} {\bibfnamefont {N.~S.}\ \bibnamefont
  {Sokolov}},\ }\href {\doibase https://doi.org/10.1016/0038-1098(82)90160-0}
  {\bibfield  {journal} {\bibinfo  {journal} {Solid State Commun.}\ }\textbf
  {\bibinfo {volume} {42}},\ \bibinfo {pages} {389} (\bibinfo {year}
  {1982})}\BibitemShut {NoStop}%
\bibitem [{\citenamefont {Harukawa}\ \emph {et~al.}(2000)\citenamefont
  {Harukawa}, \citenamefont {Murakami}, \citenamefont {Tamon}, \citenamefont
  {Ijuin}, \citenamefont {Ohmori},\ and\ \citenamefont
  {Abe}}]{SI_HARUKAWA20001231}%
  \BibitemOpen
  \bibfield  {author} {\bibinfo {author} {\bibfnamefont {N.}~\bibnamefont
  {Harukawa}}, \bibinfo {author} {\bibfnamefont {S.}~\bibnamefont {Murakami}},
  \bibinfo {author} {\bibfnamefont {S.}~\bibnamefont {Tamon}}, \bibinfo
  {author} {\bibfnamefont {S.}~\bibnamefont {Ijuin}}, \bibinfo {author}
  {\bibfnamefont {A.}~\bibnamefont {Ohmori}}, \ and\ \bibinfo {author}
  {\bibfnamefont {T.}~\bibnamefont {Abe}, \bibfnamefont {K~.and~Shigenari}},\
  }\href {\doibase https://doi.org/10.1016/S0022-2313(99)00524-4} {\bibfield
  {journal} {\bibinfo  {journal} {J. Lumin.}\ }\textbf {\bibinfo {volume}
  {87-89}},\ \bibinfo {pages} {1231} (\bibinfo {year} {2000})}\BibitemShut
  {NoStop}%
\bibitem [{\citenamefont {Garuthara}\ and\ \citenamefont
  {Siripala}(2006)}]{SI_GARUTHARA2006173}%
  \BibitemOpen
  \bibfield  {author} {\bibinfo {author} {\bibfnamefont {R.}~\bibnamefont
  {Garuthara}}\ and\ \bibinfo {author} {\bibfnamefont {W.}~\bibnamefont
  {Siripala}},\ }\href {\doibase https://doi.org/10.1016/j.jlumin.2005.11.010}
  {\bibfield  {journal} {\bibinfo  {journal} {J. Lumin.}\ }\textbf {\bibinfo
  {volume} {121}},\ \bibinfo {pages} {173} (\bibinfo {year}
  {2006})}\BibitemShut {NoStop}%
\bibitem [{\citenamefont {Koirala}\ \emph {et~al.}(2013)\citenamefont
  {Koirala}, \citenamefont {Naka},\ and\ \citenamefont
  {Tanaka}}]{SI_KOIRALA2013524}%
  \BibitemOpen
  \bibfield  {author} {\bibinfo {author} {\bibfnamefont {S.}~\bibnamefont
  {Koirala}}, \bibinfo {author} {\bibfnamefont {N.}~\bibnamefont {Naka}}, \
  and\ \bibinfo {author} {\bibfnamefont {K.}~\bibnamefont {Tanaka}},\ }\href
  {\doibase https://doi.org/10.1016/j.jlumin.2012.07.035} {\bibfield  {journal}
  {\bibinfo  {journal} {J. Lumin.}\ }\textbf {\bibinfo {volume} {134}},\
  \bibinfo {pages} {524} (\bibinfo {year} {2013})}\BibitemShut {NoStop}%
\bibitem [{\citenamefont {Koirala}\ \emph {et~al.}(2014)\citenamefont
  {Koirala}, \citenamefont {Takahata}, \citenamefont {Hazama}, \citenamefont
  {Naka},\ and\ \citenamefont {Tanaka}}]{SI_KOIRALA201465}%
  \BibitemOpen
  \bibfield  {author} {\bibinfo {author} {\bibfnamefont {S.}~\bibnamefont
  {Koirala}}, \bibinfo {author} {\bibfnamefont {M.}~\bibnamefont {Takahata}},
  \bibinfo {author} {\bibfnamefont {Y.}~\bibnamefont {Hazama}}, \bibinfo
  {author} {\bibfnamefont {N.}~\bibnamefont {Naka}}, \ and\ \bibinfo {author}
  {\bibfnamefont {K.}~\bibnamefont {Tanaka}},\ }\href {\doibase
  https://doi.org/10.1016/j.jlumin.2014.06.027} {\bibfield  {journal} {\bibinfo
   {journal} {J. Lumin.}\ }\textbf {\bibinfo {volume} {155}},\ \bibinfo {pages}
  {65} (\bibinfo {year} {2014})}\BibitemShut {NoStop}%
\bibitem [{\citenamefont {Frazer}\ \emph {et~al.}(2017)\citenamefont {Frazer},
  \citenamefont {Chang}, \citenamefont {Schaller}, \citenamefont
  {Poeppelmeier},\ and\ \citenamefont {Ketterson}}]{SI_FRAZER2017281}%
  \BibitemOpen
  \bibfield  {author} {\bibinfo {author} {\bibfnamefont {L.}~\bibnamefont
  {Frazer}}, \bibinfo {author} {\bibfnamefont {K.~B.}\ \bibnamefont {Chang}},
  \bibinfo {author} {\bibfnamefont {R.~D.}\ \bibnamefont {Schaller}}, \bibinfo
  {author} {\bibfnamefont {K.~R.}\ \bibnamefont {Poeppelmeier}}, \ and\
  \bibinfo {author} {\bibfnamefont {J.~B.}\ \bibnamefont {Ketterson}},\ }\href
  {\doibase https://doi.org/10.1016/j.jlumin.2016.11.011} {\bibfield  {journal}
  {\bibinfo  {journal} {J. Lumin.}\ }\textbf {\bibinfo {volume} {183}},\
  \bibinfo {pages} {281} (\bibinfo {year} {2017})}\BibitemShut {NoStop}%
\bibitem [{\citenamefont {Nolan}(2008)}]{SI_Nolan008}%
  \BibitemOpen
  \bibfield  {author} {\bibinfo {author} {\bibfnamefont {M.}~\bibnamefont
  {Nolan}},\ }\href {\doibase https://doi.org/10.1016/j.tsf.2008.04.020}
  {\bibfield  {journal} {\bibinfo  {journal} {Thin Solid Films}\ }\textbf
  {\bibinfo {volume} {516}},\ \bibinfo {pages} {8130} (\bibinfo {year}
  {2008})}\BibitemShut {NoStop}%
\bibitem [{\citenamefont {Scanlon}\ \emph
  {et~al.}(2009{\natexlab{b}})\citenamefont {Scanlon}, \citenamefont {Morgan},\
  and\ \citenamefont {Watson}}]{SI_ScanlonJCP2009}%
  \BibitemOpen
  \bibfield  {author} {\bibinfo {author} {\bibfnamefont {D.~O.}\ \bibnamefont
  {Scanlon}}, \bibinfo {author} {\bibfnamefont {B.~J.}\ \bibnamefont {Morgan}},
  \ and\ \bibinfo {author} {\bibfnamefont {G.~W.}\ \bibnamefont {Watson}},\
  }\href {\doibase 10.1063/1.3231869} {\bibfield  {journal} {\bibinfo
  {journal} {J. Chem. Phys.}\ }\textbf {\bibinfo {volume} {131}},\ \bibinfo
  {pages} {124703} (\bibinfo {year} {2009}{\natexlab{b}})}\BibitemShut
  {NoStop}%
\bibitem [{\citenamefont {Huang}(2016)}]{SI_Huang2016}%
  \BibitemOpen
  \bibfield  {author} {\bibinfo {author} {\bibfnamefont {B.}~\bibnamefont
  {Huang}},\ }\href {\doibase https://doi.org/10.1016/j.ssc.2016.01.008}
  {\bibfield  {journal} {\bibinfo  {journal} {Solid State Commun.}\ }\textbf
  {\bibinfo {volume} {230}},\ \bibinfo {pages} {49} (\bibinfo {year}
  {2016})}\BibitemShut {NoStop}%
\bibitem [{\citenamefont {Isseroff}\ and\ \citenamefont
  {Carter}(2013)}]{SI_Isserof2013}%
  \BibitemOpen
  \bibfield  {author} {\bibinfo {author} {\bibfnamefont {L.~Y.}\ \bibnamefont
  {Isseroff}}\ and\ \bibinfo {author} {\bibfnamefont {E.~A.}\ \bibnamefont
  {Carter}},\ }\href {\doibase 10.1021/cm3040278} {\bibfield  {journal}
  {\bibinfo  {journal} {Chem. Mater.}\ }\textbf {\bibinfo {volume} {25}},\
  \bibinfo {pages} {253} (\bibinfo {year} {2013})}\BibitemShut {NoStop}%
\bibitem [{\citenamefont {Kr{\"o}ger}\ and\ \citenamefont
  {Vink}(1956)}]{SI_KROGER1956307}%
  \BibitemOpen
  \bibfield  {author} {\bibinfo {author} {\bibfnamefont {F.~A.}\ \bibnamefont
  {Kr{\"o}ger}}\ and\ \bibinfo {author} {\bibfnamefont {H.~J.}\ \bibnamefont
  {Vink}},\ }\href {\doibase 10.1016/S0081-1947(08)60135-6} {\bibfield
  {journal} {\bibinfo  {journal} {Solid State Phys.}\ }\textbf {\bibinfo
  {volume} {3}},\ \bibinfo {pages} {307} (\bibinfo {year} {1956})}\BibitemShut
  {NoStop}%
\end{thebibliography}%


\clearpage
\clearpage 
\setcounter{page}{1}
\renewcommand{\thetable}{S\arabic{table}} 
\setcounter{table}{0}
\renewcommand{\thefigure}{S\arabic{figure}}
\setcounter{figure}{0}
\renewcommand{\thesection}{S\arabic{section}}
\setcounter{section}{0}
\renewcommand{\theequation}{S\arabic{equation}}
\setcounter{equation}{0}
\onecolumngrid

\begin{center}
\textbf{Supplementary information for\\\vspace{0.5 cm}
\large Importance of surface oxygen vacancies for ultrafast hot carrier relaxation and transport in \ce{Cu2O}\\\vspace{0.3 cm}}
Chiara Ricca,$^{1, 2}$, Lisa Grad$^{3}$, Matthias Hengsberger$^{3}$, J{\"u}rg Osterwalder$^{3}$, and Ulrich Aschauer$^{1, 2}$

\small
$^1$\textit{Department of Chemistry and Biochemistry, University of Bern, Freiestrasse 3, CH-3012 Bern, Switzerland}

$^2$\textit{National Centre for Computational Design and Discovery of Novel Materials (MARVEL), Switzerland}

$^3$\textit{Department of Physics, University of Zurich, Winterthurerstrasse 190, CH-8057 Zurich, Switzerland}

(Dated: \today)
\end{center}

\section{Method details}

\subsection{\label{sec:compbulk}Bulk calculations}

\ce{Cu2O} (cuprous oxide, mineral cuprite) has a cubic structure with four Cu and two O atoms in the unit cell (space group $Pn\bar{3}m$). A $2\times2\times2$ supercell of the 6-atom unit cell was used to model oxygen-deficient bulk cuprous oxide with a $\Gamma$-centered $6\times6\times6$ Monkhorst-Pack \citeSI{SI_monkhorst1976special} \textbf{k}-point grid to sample the Brillouin zone. Defects were created by removing one oxygen atom (V$_{\textrm{O}}$, concentration $\approx$ 6\%) from this 48-atom supercell. For the stoichiometric bulk, both lattice parameters and atomic positions were allowed to relax, while for defect and surface calculations the lattice parameters are kept fixed at their relaxed bulk values. Structural relaxations were performed until forces converged below 10$^{-3}$~eV/\AA\ and stress components below 5$\times$10$^{-5}$~eV/\AA$^3$. Electrostatic corrections~\citeSI{SI_makov1995periodic, SI_lany2009accurate} were considered for bulk defects using the experimental value of 7.1~\citeSI{SI_Stolz2018} for the dielectric constant.

\subsection{\label{sec:compchempot}\ce{Cu2O} chemical potentials}

When calculating oxygen vacancy formation energies, different synthesis conditions can be accounted for by adjusting the oxygen chemical potential ($\mathrm{\mu_O = \mu_O^0 + \Delta \mu_O}$) assuming that it is in equilibrium with a given oxygen reservoir. If $\mu_\textrm{O} = \frac{1}{2}\mu(\textrm{O}_2)+ \Delta \mu_{\textrm{O}}$, where $\mu(\textrm{O}_2)$ is the total energy of \ce{O2} in its triplet state, then $\Delta \mu_{\textrm{O}}$ can vary within a range limited by the formation of CuO for the oxygen-rich limit and by the decomposition of \ce{Cu2O} to metallic Cu for the oxygen-poor limit. The first condition is given by
\begin{align}
	\Delta\mu_{\ce{Cu}} + \Delta\mu_{\ce{O}} = \Delta H_f^{\ce{CuO}} = -1.48~\mathrm{eV},
\end{align}
where $\Delta H_f^{\ce{CuO}}$ is the heat of formation of CuO computed at the HSE level of theory. The second condition is given by
\begin{align}
	\mathrm{\mu_{Cu} = \mu_{Cu;metal}} \rightarrow \Delta \mu_{\mathrm{Cu}} = 0
\end{align}
and since the chemical potentials are related by
\begin{align}
	2 \Delta \mu_{\ce{Cu}}+\Delta\mu_{\ce{O}} = \Delta H_f^{\ce{Cu2O}} = -1.55~\mathrm{eV},
\end{align}
where $\Delta H_f^{\ce{Cu2O}}$ is the heat of formation of cuprous oxide computed with the HSE functional, this results in
\begin{align}
	\Delta \mu_{\ce{O}} &= -1.41~\mathrm{eV} \\
	\Delta \mu_{\ce{Cu}} &= -0.07~\mathrm{eV}
\end{align}
and 
\begin{align}
	\Delta \mu_{\ce{O}}= -1.55~\mathrm{eV}
	\Delta \mu_{\ce{Cu}}= 0.00~\mathrm{eV}
\end{align}
for the O-rich and O-poor case, respectively.

\subsection{\label{sec:compcapture}Non-radiative carrier capture}

We considered non-radiative carrier capture processes involving a defect and occurring via multi-phonon emission~\citeSI{SI_Huang1950, SI_HenryLang1977}. In such a process, a system in an initial excited state ($i$), for example a neutral V$_\mathrm{O}^{\bullet \bullet}$ in \ce{Cu2O} with a hole ($h^+$) in the valence band, vibrates around its equilibrium geometry. Because of electron-phonon coupling, the deformation of the structure induces oscillations in the electronic energy level of the defect state, eventually allowing hole capture, \textit{i.e.} an energy-conserving transition to the final electronic ground state ($f$) constituted by the singly positively charged V$_\mathrm{O}^{\bullet}$, followed by the relaxation of the system towards the equilibrium geometry of the ground state by emitting multiple phonons. Such a process can be described by first principles using the approach introduced by Alkauskas \textit{et al.}~\citeSI{SI_AlkauskasPRB2014,SI_AlkauskasPRB2016} based on the static approximation~\citeSI{SI_Huang1981} and using an effective one-dimensional configuration coordinate $Q$ to represent the phonon wavefunction describing all the vibrations coupling to the change of the defect's geometry caused by the carrier capture:
\begin{equation}
Q^2=\sum_{\alpha,i}m_\alpha \Delta R_{\alpha,i}^2,
\label{eq:Q}
\end{equation}
where $m_\alpha$ and $\Delta R_{\alpha,i}$ are the mass and the displacement of atom $\alpha$ from its equilibrium position in one of the two charge states along the $i$-direction. Within this approach, the capture coefficient is given by
\begin{equation}
C(T)=sV \eta_{sp} g \frac{2\pi }{\hbar}W_{if}^2 \sum_{m,n} w_m(T) \vert\langle \chi_{im} \vert Q + \Delta Q \vert \chi_{fn} \rangle \vert ^2 
\times \delta( \Delta E + \epsilon_{im} -\epsilon_{fn}) ,
\label{eq:capturecoeffSI}
\end{equation}
where $V$ is the volume of the supercell, $\eta_{sp} $ accounts for spin-selection rules ($\eta_{sp}=1/2 $ when the initial state is a spin doublet and the final state is a spin singlet), $g$ is the degeneracy of the final state, $W_{if}$ is the electron-phonon coupling matrix element of the initial and final states, $\Delta E$ is the energy difference between the two states, $\chi$ and $\epsilon$ are the phonon wave functions and eigenvalues, respectively, for the excited ($im$) and ground ($fn$) electronic states, while $w_m$ is the thermal occupation number of the excited vibrational state. $\delta$ is replaced with a Gaussian function with a finite width. Matrix elements $W_{if}$ are computed according to Ref.~\citeSI{SI_AlkauskasPRB2014}. The Sommerfeld factor $s$, computed as described in Ref.~\citeSI{SI_KimJmatChemA2019, SI_KimRSC2020}, is necessary to describe the capture by charged defects since it accounts for the Coulomb interaction at a temperature $T$ between a carrier with charge $q$ and a charged defect in a charge state $Q$:
\begin{equation}
s=
 \begin{cases}
 4\left|Z\right|( \pi E_\mathrm{R}/k_\mathrm{B}T)^{1/2} & \text{for $Z < $0} \\
 8/\sqrt{3}( \pi^2 Z^2 E_\mathrm{R}/k_\mathrm{B}T)^{2/3} exp(-3(Z^2\pi^2E_\mathrm{R}/k_\mathrm{B}T)^{1/3}) & \text{for $Z > $0}
 \end{cases}
\label{eq:sommerfeld}
\end{equation}
where $Z= Q/q$ is negative for an attractive and positive for a repulsive center, $k_\mathrm{B}$ is the Boltzmann constant, $E_\mathrm{R} = m^*q^4/(2 \hbar^2 \epsilon^2)$, with $\epsilon$ the dielectric constant of \ce{Cu2O} and $m^*$ the carrier effective mass. 

Carrier capture cross-sections ($\sigma$) are derived from $C(T)=\sigma(T) v(T)$, where $v$ is the carrier thermal velocity $v (T)= \sqrt{3k_\mathrm{B}T/m^*}$.

\newpage
\section{\label{SISec:defbulk}Stoichiometric bulk $\mathbf{Cu_{2}O}$}

\begin{figure}[h]
 \centering
 \includegraphics[width=0.3\columnwidth]{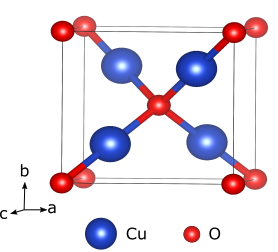}
 \caption{$Pn\bar{3}m$ unit cell of \ce{Cu2O}.}
\label{fig:Cu2Ounitcell}
\end{figure}

\ce{Cu2O} crystallizes in a simple cubic lattice (space group $Pn\bar{3}m$) with six atoms in the unit cell: four Cu positioned on a face-centered cubic lattice and the two O atoms occupying the tetrahedral sites and forming a body-centered cubic sub-lattice (see Fig.~\ref{fig:Cu2Ounitcell}). Consequently, the copper atoms are linearly coordinated by two nearest neighbor O atoms ($D_{3h}$ symmetry), while each oxygen is fourfold coordinated ($T_d$ symmetry). As can be seen from Table~\ref{tbl:prop_stoic}, the structural parameters computed with the hybrid HSE functional are in excellent agreement with experiment (a mean absolute error of 0.4\% on the lattice parameters and the Cu--O distance), in line with previous reports~\citeSI{SI_ScanlonPRL2009, SI_ScanlonJPCL2010, SI_Heinemann2013}. In line with previous studies based on hybrid DFT~\citeSI{SI_Nolan2006, SI_ScanlonPRL2009, SI_ScanlonJPCL2010, SI_Heinemann2013}, our predicted band gap is in excellent agreement with experiment, being underestimated by only 0.17 eV (see Table~\ref{tbl:prop_stoic}).
\begin{table}[h]
\caption{Comparison of the calculated and experimental
structural properties (lattice parameters $a$ in \AA, Cu--O distance d in \AA, and band gap $E_\mathrm{g}$ in eV).}
\begin{tabular*}{\columnwidth}{@{\extracolsep{\fill}}lcc}
\hline
\hline
	& HSE & Expt.\\
\hline
$a$ & 4.288 & 4.270~\citeSI{SI_Werner1982}\\
d & 1.857 & 1.850~\citeSI{SI_Werner1982}\\
$E_\mathrm{g}$ & 2.00 & 2.17\citeSI{SI_Lide2008}\\
\hline
\end{tabular*}
\label{tbl:prop_stoic}
\end {table}

\begin{figure}[h]
 \centering
 \includegraphics[width=0.4\columnwidth]{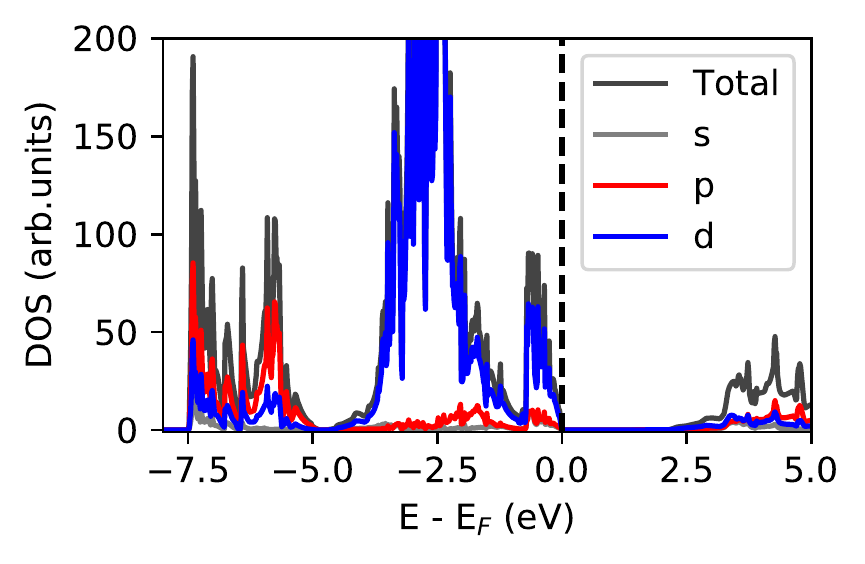}
 \caption{ Total and orbital-projected density of states (DOS and
PDOS, respectively) for bulk \ce{Cu2O} computed at the HSE level of theory. The
zero of the energy scale was set at the Fermi energy.}
\label{fig:PDOS_stoic}
\end{figure}

\begin{figure}[h]
 \centering
 \includegraphics[width=0.3\columnwidth]{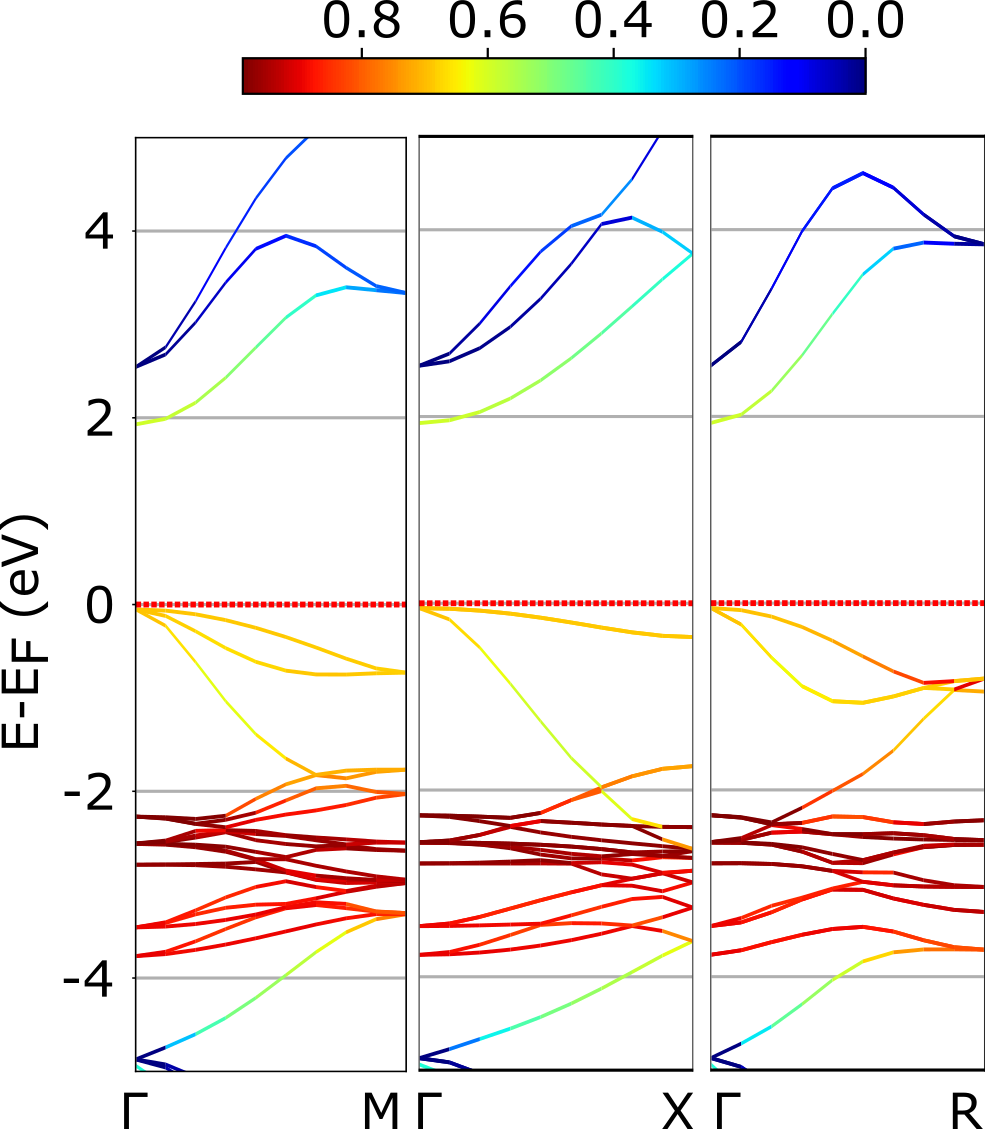}
 \caption{Electronic bandstructure of \ce{Cu2O} computed at the HSE level. The contribution of the Cu-$3d$ to the different bands is shown as coloured fat bands. }
\label{fig:bandstructure}
\end{figure}

Figures~\ref{fig:PDOS_stoic} and ~\ref{fig:bandstructure} shows the predicted electronic structure of \ce{Cu2O} with the highest lying valence band (VB) states mainly formed by the antibonding Cu-$3d$ states with a small admixing of O-$2p$ states mainly contributing at lower energies in the VB. The bottom of the conduction is formed, instead, by Cu-$4s$ and $3d$ states. The experimentally observed $p$-type behavior of cuprous oxide is a consequence of this peculiar electronic structure with the formally fully occupied Cu-$3d$ states at the top of the VB, as opposed to the majority of oxides that show O-$2p$ character for the highest lying VB states. Additionally, for a good $p$-type conductor, the VB should be highly dispersive leading to a low hole effective mass and thus high mobility~\citeSI{SI_Raebiger2007} (see Fig.~\ref{fig:bandstructure}).

The electron and hole effective masses computed with HSE for \ce{Cu2O} along the high symmetry directions at the band extrema at $\Gamma$ are reported in Table~\ref{tbl:effectivemasses}. In absence of spin-orbit splitting, the states at the VBM at $\Gamma$ are triply degenerate and we can identify two heavy and one light hole band. The effective masses are isotropic, in line with experiments~\citeSI{SI_Ohyama1997}. By averaging over the three degenerate bands, we obtain a conduction effective mass for the hole of 0.54$m_0$. Our calculated effective masses are slightly smaller than the available experimental values. Our results, however, are different from the theoretical prediction of Nie \textit{et al.}~\citeSI{SI_Nie2002} who, while using only LDA, included spin-orbit interaction, which, mixing the wavefunctions at the top of VB, has a very large effect on the calculated effective masses and removes the degeneracy of the VBM, resulting in the light hole state becoming the top of the VB.
\begin{table}[h]
\caption{Calculated electron and hole effective masses (in units of free electron mass $m_0$) at $\Gamma$. Results are compared with available experimental data.}
\begin{tabular*}{\columnwidth}{@{\extracolsep{\fill}}lccccc}
\hline
\hline
	& \multicolumn{4}{c}{HSE} & Expt.\\
	& [100] & [010] & [111] &Average & \\
\hline
CBM & 0.83 & 0.83 & 0.83 & 0.83 &0.99~\citeSI{SI_Hodby_1976},0.98~\citeSI{SI_GOLTZENE1976}\\
\hline
 & 0.20 & 0.20 & 0.20 & &0.58~\citeSI{SI_Hodby_1976},0.69~\citeSI{SI_GOLTZENE1976}\\
VBM & 3.89 & 3.88 &3.88& 0.54 &\\
 & 3.89 & 3.89 & 3.88 & &\\
\hline
\hline
\end{tabular*}
\label{tbl:effectivemasses}
\end {table}

\newpage
\section{\label{sec:defbulk}Defective bulk $\mathbf{Cu_{2}O}$}

The properties of oxygen vacancies (V$_\mathrm{O}$) in cuprous oxide are still widely debated with contradictory results reported by both experiments~\citeSI{SI_Bloem1958, SI_ZOUAGHI1969311, SI_GASTEV1982389, SI_HARUKAWA20001231, SI_GARUTHARA2006173, SI_KOIRALA2013524, SI_KOIRALA201465, SI_FRAZER2017281} and theory~\citeSI{SI_Nolan2006, SI_Raebiger2007, SI_Nolan008, SI_ScanlonJCP2009, SI_ScanlonJPCL2010, SI_ScanlonPRL2009, SI_Huang2016}. \ce{Cu2O} predominantly has Schottky defects (O and Cu vacancy pairs) in concentrations determined by the synthesis temperature. Donor states in \ce{Cu2O} samples synthesized at oxygen partial pressures of $10^{-3}-10^{-1}$ mmHg were reported for the first time in photoluminescence (PL) experiments~\citeSI{SI_Bloem1958}, and attributed to singly and doubly ionized V$_\mathrm{O}$. A donor level 0.38~eV below the CB minimum (CBM) and attributed to V$_\mathrm{O}$ was subsequently confirmed in PL spectra of electrodeposited cuprous oxide by different authors~\citeSI{SI_ZOUAGHI1969311, SI_GASTEV1982389, SI_HARUKAWA20001231, SI_GARUTHARA2006173}. In more recent PL studies, the broad luminescence bands at 1.72 eV and 1.53~eV from the VB were assigned to doubly and singly positively charged V$_\mathrm{O}$, respectively~\citeSI{SI_KOIRALA2013524, SI_KOIRALA201465, SI_FRAZER2017281}, even though the exact position of the peaks and their relative intensity was found to strongly depend on the synthesis and post-processing conditions. 

Early DFT results~\citeSI{SI_Nolan2006} reported that under oxidizing conditions ($E_\mathrm{F}=0$ in Eq.~\ref{eq:formenerg} of the main text) the most stable V$_\mathrm{O}$ is in the +2 charge state. This result was obtained using the general gradient approximation (GGA) PBE exchange-correlation functional. A subsequent study, reported, however, a completely different behavior using the same functional: no donor level associated with V$_\mathrm{O}$ was observed inside the band gap, suggesting that V$_\mathrm{O}$ cannot act as a hole killer in cuprous oxide~\citeSI{SI_Raebiger2007}. Similar conclusions were reached with the more accurate hybrid HSE functional~\citeSI{SI_ScanlonJCP2009, SI_ScanlonJPCL2010, SI_ScanlonPRL2009}. Discrepancies of theoretical results can be explained by different DFT functionals, but also by different corrections for charged defects in periodic supercells, and by computational details like the sampling of the Brillouin zone. The validity of earlier results~\citeSI{SI_ScanlonJCP2009, SI_ScanlonJPCL2010, SI_ScanlonPRL2009} obtained using a coarser \textbf{k}-point sampling (presumably due to the high cost of HSE calculations and computational limitations at that time) was shown to lead to incorrect predictions for Cu vacancies in \ce{Cu2O}~\citeSI{SI_Isserof2013}. Given that more accurate calculations are now affordable, we repeated the study of V$_\mathrm{O}$ in \ce{Cu2O} using the HSE functional but with a denser \textbf{k}-point mesh.

\begin{figure}[h]
 \centering
 \includegraphics[width=0.4\columnwidth]{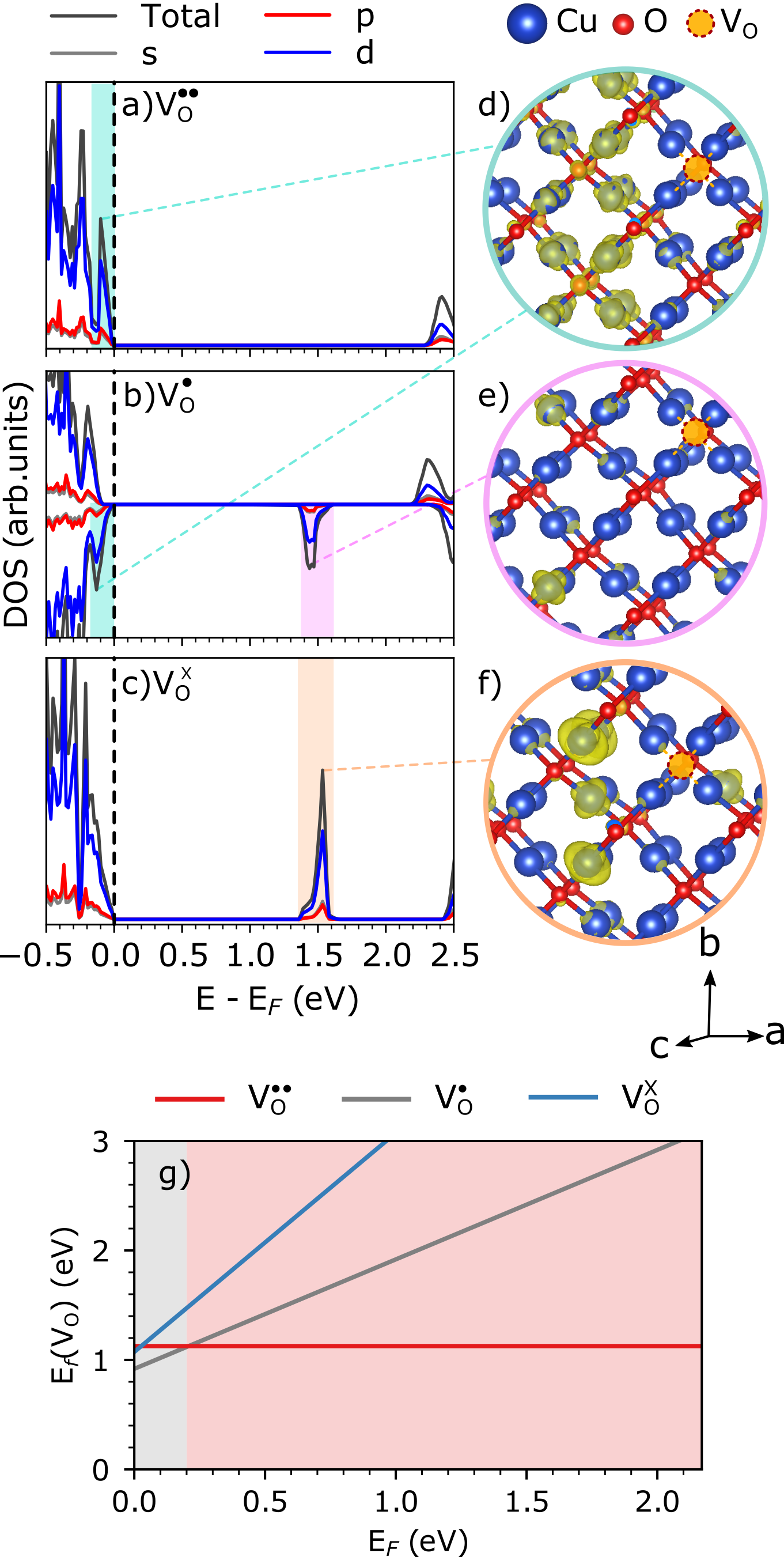}
 \caption{Total and projected density of states for one a) V$_\mathrm{O}^{\bullet\bullet}$, b) V$_\mathrm{O}^{\bullet}$, and c) V$\mathrm{_O^\mathrm{X}}$ in bulk \ce{Cu2O}. The zero of the energy scale is at the Fermi energy. For the spin-polarized V$_\mathrm{O}^{\bullet}$ calculation, the DOS for the spin-up and spin-down channels are reported with positive and negative values along the $y$-axis. The isosurfaces (2$\times 10 ^{-2}$ e/\AA$^3$) in d)-f) show the charge density associated with the defect states highlighted with the corresponding color in panels a)-c). g) Oxygen vacancy formation energy ($E_f(\mathrm{V_O})$) under O-poor conditions in different charge states as a function of the Fermi energy ranging from the valence band maximum ($E_\mathrm{F}=0$) up to the experimental band gap for stoichiometric bulk \ce{Cu2O}.}
\label{fig:PDOS_bulk_VO}
\end{figure}

Figure~\ref{fig:PDOS_bulk_VO} shows the computed density of states for a neutral, singly, and doubly positively charged V$_\mathrm{O}$ in bulk \ce{Cu2O}. These are designated in Kr{\"o}ger-Vink notation~\citeSI{SI_KROGER1956307} as V$_\mathrm{O}^{\bullet\bullet}$, V$_\mathrm{O}^{\bullet}$, and V$\mathrm{_O^\mathrm{X}}$ respectively, where $\bullet$ and X superscripts indicate, respectively, a charge of +1 and 0 relative to the respective lattice site, $h^+/e^-$ represent free holes/electrons, and a transition from V$_\mathrm{O}^{\bullet}+ e^-$ to V$_\mathrm{O}^{\bullet \bullet}$ designates the electron capture process by a V$_\mathrm{O}^{\bullet}$ defect.  The two extra electrons left in the lattice upon V$_\mathrm{O}^{\bullet\bullet}$ formation are resonant in the VB (see Fig.~\ref{fig:PDOS_bulk_VO}a and d). If one oxygen atom and one electron are simultaneously removed to form V$_\mathrm{O}^{\bullet}$, the defect state is split in a filled state still merged with the top of VB and occupied by the remaining extra electron, and an unoccupied state lying at 1.45~eV in the gap and mainly localized on two Cu atoms far from the V$_\mathrm{O}$ (see Fig.~\ref{fig:PDOS_bulk_VO}b and e). Finally, V$\mathrm{_O^\mathrm{X}}$ is associated with an empty defect state at about 1.53~eV localized on 4 Cu atoms in next nearest neighbor positions to the defect (see Fig.~\ref{fig:PDOS_bulk_VO}c and f). Structural relaxations observed upon defect formation are minimal and restricted to the four Cu atoms in nearest neighbor positions to V$_\mathrm{O}$ that move towards the vacancy, thus elongating their remaining Cu--O bonds by about 0.13~\AA\ for V$_\mathrm{O}^{\bullet\bullet}$ and V$\mathrm{_O^\mathrm{X}}$ and 0.14~\AA\ for V$_\mathrm{O}^{\bullet}$.

Figure~\ref{fig:PDOS_bulk_VO}g illustrates the changes in the formation energies for a V$_\mathrm{O}$ in the considered charge states as a function of the position of the Fermi energy under O-poor synthesis conditions. In agreement with previous results~\citeSI{SI_Raebiger2007, SI_ScanlonJCP2009, SI_ScanlonJPCL2010, SI_ScanlonPRL2009}, V$_\mathrm{O}^{\bullet\bullet}$ has a relatively low formation energy. However, we observe that for Fermi energies very close to the top of the valence band V$_\mathrm{O}^{\bullet}$ becomes the favored charge state. Thus, contrarily to previous results~\citeSI{SI_ScanlonJCP2009, SI_ScanlonJPCL2010, SI_ScanlonPRL2009} with the same hybrid DFT functional, but with a coarser \textbf{k}-point grid, the computed thermodynamic transition level $\epsilon(+1/0)$ at 0.21~eV suggests that, under oxidizing conditions ($E_\mathrm{F}=0$ in eq.~\ref{eq:formenerg}), V$_\mathrm{O}$ in \ce{Cu2O} could potentially act as hole traps, due to the proximity of the $\epsilon(+1/0)$ transition level to the VB , and hence to the low ionization energy for the V$_\mathrm{O}^{\bullet}$/V$_\mathrm{O}^{\bullet\bullet}$  couple.

\clearpage
\section{\label{sec:capture}Carrier Capture}
\begin{figure}[h]
	\centering
	\includegraphics[width=0.5\columnwidth]{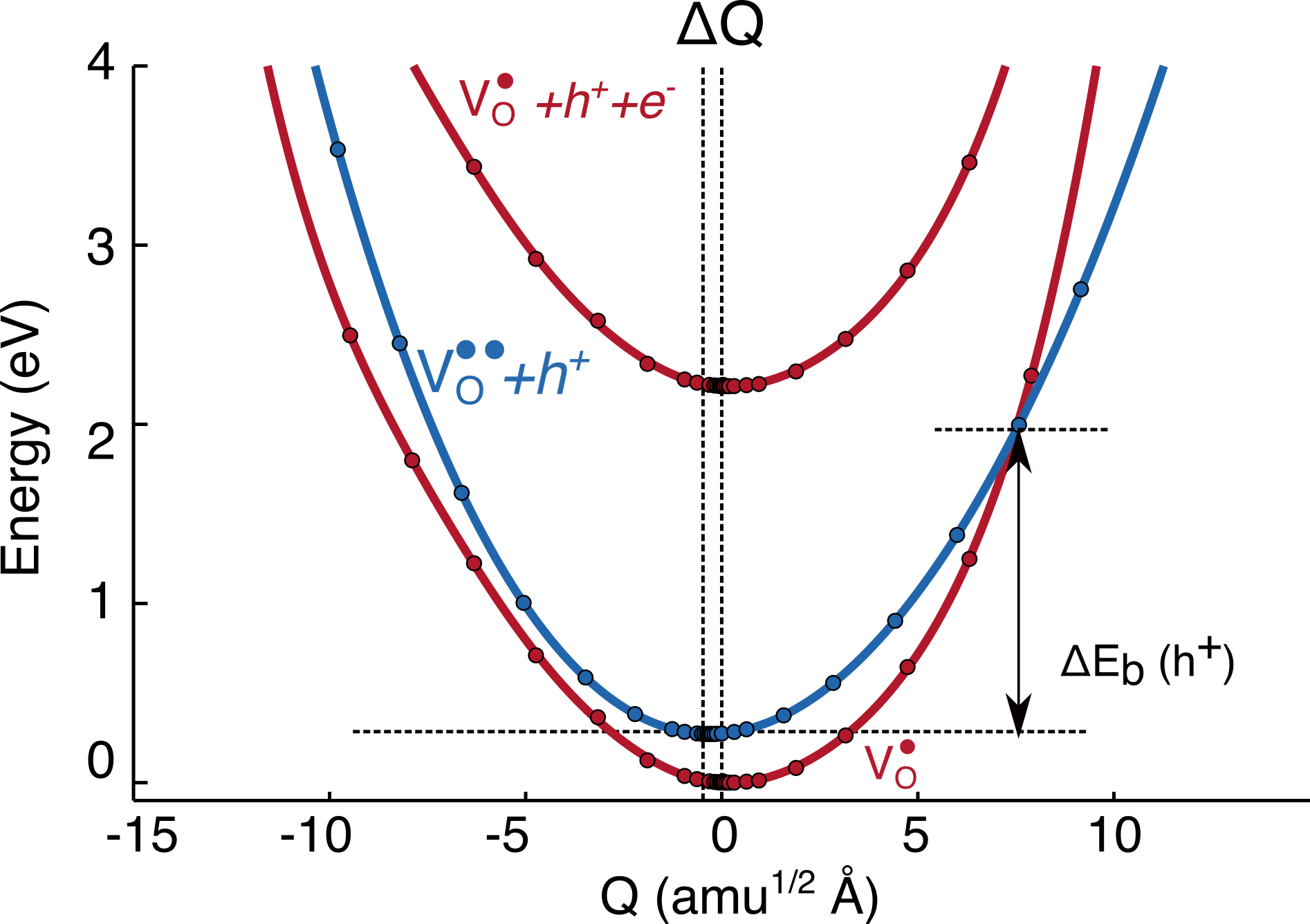}
	\caption{Configuration coordinate diagram for carrier capture a) for V$_\mathrm{O}$(+/0) in bulk \ce{Cu2O} and b) for V$_\mathrm{O}$(+2/+1) at the reconstructed \ce{Cu2O}-(111) surface. The solid circles represent the relative formation energies calculated using hybrid DFT and the lines are spline fits. $\Delta$E$_\mathrm{b}$ is the carrier capture barrier.}
	\label{fig:carriercapture_confcord}
\end{figure}

\begin{figure}[h]
 \centering
 \includegraphics{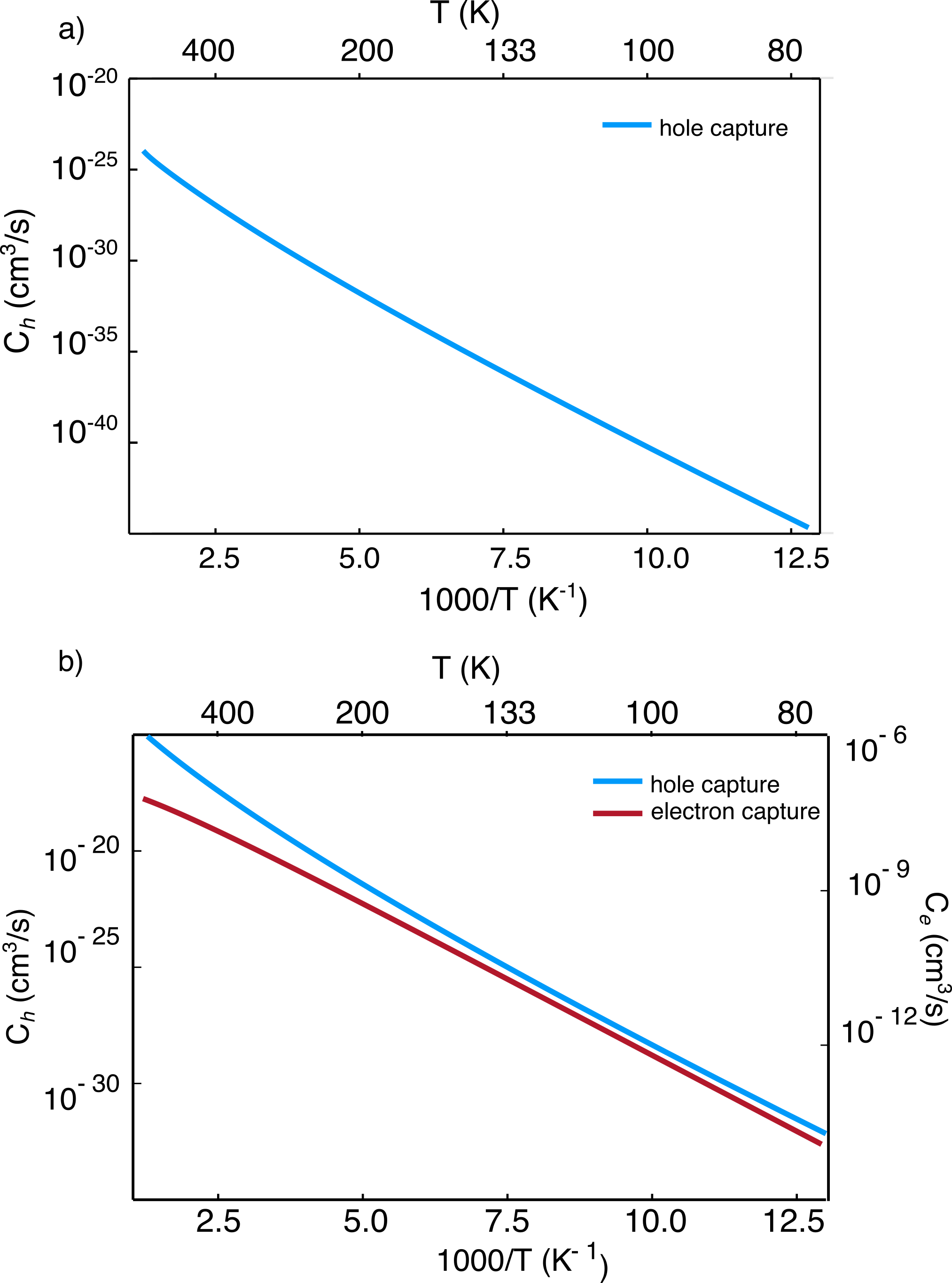}
 \caption{Temperature dependence of the carrier capture coefficients computed for a) bulk \ce{Cu2O} or b) for the \ce{Cu2O} (111) surface.}
\label{fig:carriercapture_vs_T}
\end{figure}

The calculated 1D configuration coordinate diagram for carrier capture in bulk \ce{Cu2O} is shown in Fig.~\ref{fig:carriercapture_confcord} with extracted key parameters in Table~\ref{tbl:carriercaptureSI}. The ground state is the V$_\mathrm{O}^{\bullet}$ defect, while the excited state corresponds to the neutral oxygen vacancy and a hole in the VB (V$_\mathrm{O}^{\bullet \bullet }+h^+$). The separation between the minima of the two potential energy curves is $\Delta E=0.21~\mathrm{eV}$, corresponding to the charge transition level with respect to the VBM. The minima are offset horizontally by only $\Delta Q = 0.31~\mathrm{amu^{\sfrac{1}{2}}\AA}$, in line with the minimal structural relaxations upon V$_\mathrm{O}$ formation in bulk \ce{Cu2O}. As a consequence, a sizable hole capture barrier $\Delta E_b = 1.685~\mathrm{eV}$ leads to a small hole capture coefficient $C = 1.04\times10^{-28}~\mathrm{cm^3/s}$ and hole capture cross-section $\sigma = 6.57\times 10^{-20}~\mathrm{\AA^2}$ at 298~K (for temperature-dependent values see SI Fig.~\ref{fig:carriercapture_vs_T}). Electron capture, on the other hand, cannot happen since the V$_\mathrm{O}^{\bullet \bullet }+h^+$ and V$_\mathrm{O}^{\bullet }+h^+ + e^-$ curves in Fig.~\ref{fig:carriercapture_confcord}a do not intersect.

Table \ref{tbl:carriercaptureSI} reports additional key parameters involved in the carrier capture process compared to the table in the main text.
\begin{table}[h]
\caption{Key parameters for the carrier capture process in the oxygen-deficient bulk cuprous oxide or at the reconstructed \ce{Cu2O}-(111) surface: degeneracy factor $g$ of the final state, total mass-weighted distortions ($\Delta Q$, in $\mathrm{amu^{\sfrac{1}{2}}\AA}$), ionization energy ($\Delta E$, in eV), carrier capture barrier ($\Delta E_b$, in eV), electron-phonon coupling matrix element ($W_{if}$, in eV/amu$^{1/2}$\AA), and Sommerfeld factor ($s$), carrier capture coefficient ($C$, in cm$^3$/s) and carrier capture cross section ($\sigma$, \AA$^2$) at 298 K.}
\begin{tabular*}{\columnwidth}{@{\extracolsep{\fill}}lcccccccccc}
\hline
\hline
System & Defect & Carrier & $g$ & $\Delta Q$ &$\Delta E$ 	& $\Delta E_b$ & $W_{if}$ & $s$ & $C$ &$\sigma$ \\
\hline
Bulk & (+/0) & holes & 2 & 0.31 & 0.21 & 1.685 & 0.00039 & 7.2 $\times$ 10$^{-4}$ & 1.04$\times$10$^{-28}$ & 6.57 $\times$10$^{-20}$ \\
\hline
Surface & (+2/+1) & holes & 3 & 4.83 & 0.20 & 0.003 & 0.00257 & 2.08 $\times$10$^{-6}$ & 1.24$\times$10$^{-19}$ & 7.87$\times$10$^{-11}$ \\
 & (+2/+1*) & electrons & 3 & 4.83 & 2.00 & 0.011 & 0.00206 & 33.77 & 4.87$\times$10$^{-9}$ & 3.82 \\
\hline
\hline
\end{tabular*}
\label{tbl:carriercaptureSI}
\end {table}

\newpage
Figure \ref{fig:carriercapture_vs_T} reports the temperature dependent carrier capture coefficients in the bulk and at the reconstructed (111) surface.
%
%
\bibliographystyleSI{apsrev4-1}
\bibliographySI{references}

\end{document}